\documentclass[conference,letterpaper]{IEEEtran}
\IEEEoverridecommandlockouts

\usepackage[utf8]{inputenc} 
\usepackage[T1]{fontenc}
\usepackage{url}
\usepackage{ifthen}
\usepackage{cite}
\usepackage[cmex10]{amsmath} 

\usepackage{amssymb}
\usepackage{mathtools}

\usepackage{subfig}

\newtheorem{theorem}{Theorem}[section]
\newtheorem{proposition}[theorem]{Proposition}
\newtheorem{lemma}[theorem]{Lemma}

\newtheorem{definition}[theorem]{Definition}

\newtheorem{claim}[theorem]{Claim}

\usepackage[disable]{todonotes}

\begin{document}
\title{Remote Channel Synthesis\\
} 

\author{%
  \IEEEauthorblockN{Yassine Hamdi\IEEEauthorrefmark{1}, Deniz G\"{u}nd\"{u}z\IEEEauthorrefmark{1}}
  \IEEEauthorblockA{\IEEEauthorrefmark{1}Department of Electrical and Electronic Engineering, Imperial College London, UK,}
  \IEEEauthorblockA{\{y.hamdi, d.gunduz\}@imperial.ac.uk}
}

\maketitle


\begin{abstract}
We consider the problem of synthesizing a memoryless channel between an unobserved source and a remote terminal.
An encoder has access to a partial or noisy version $Z^n = (Z_1, \ldots, Z_n)$ of a remote source sequence $X^n = (X_1, \ldots, X_n),$ with $(X_i,Z_i)$
independent and identically distributed
with joint distribution $q_{X,Z}.$
The encoder communicates
through a noiseless link to a decoder which aims to produce an output $Y^n$ coordinated with the remote source;
that is,
the total variation distance between
the joint distribution of $X^n$ and $Y^n$
and
some i.i.d. target distribution $q_{X,Y}^{\otimes n}$
is required to vanish as $n$ goes to infinity.
The two terminals may have access to a source of rate-limited common randomness.
We present a single-letter characterization of the optimal compression and common randomness rates.
We
also
show that
when
the common randomness rate
is small,
then in most cases,
coordinating $Z^n$ and $Y^n$ using a standard channel synthesis scheme is strictly sub-optimal.
In other words, schemes for which the joint distribution of $Z^n$ and $Y^n$ approaches a product distribution asymptotically are strictly sub-optimal.
\end{abstract}

\section{Introduction}

\begin{figure}
    \centering
    \includegraphics[width=0.99\columnwidth]{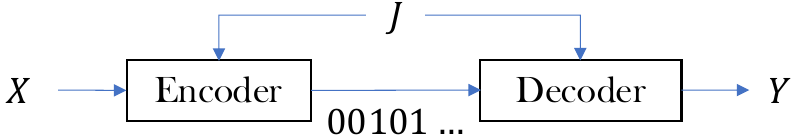}
    \caption{One shot channel synthesis with common randomness.}
    \label{fig:one_shot_channel_simulation}
\end{figure}

\begin{figure}
    \centering
    \includegraphics[width=0.99\columnwidth]{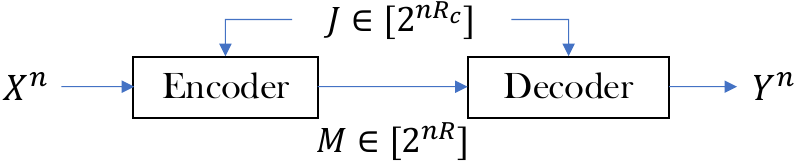}
    \caption{Block coding setup for channel synthesis, a.k.a. the reverse channel coding problem.}
    \label{fig:block_coding_channel_synthesis}
\end{figure}

The problem of one-shot channel synthesis is depicted in Figure \ref{fig:one_shot_channel_simulation}.
An encoder observes a source $X,$ and sends information to a decoder, which produces an output $Y.$ The terminals may have access to a common randomness $J.$ The goal is for the distribution of $Y$ conditioned on the realization $x$ of $X$ to match a prescribed distribution $q_{Y|X{=}x}.$
The
name \textit{relative entropy coding} is also used for a similar formulation in \cite{2020GregFlamishNeuripsFirstDefinitionRelativeEntropyCoding}.
Taking a source consisting of a sequence $X^n = (X_1, ..., X_n)$ of i.i.d. symbols of distribution $p_X$ gives
the block coding version of channel synthesis, called
\textit{the reverse channel coding problem},
or \textit{strong coordination}
\cite{2002BennettReverseShannonTheorem,2013PaulCuffDistributedChannelSynthesis,2014BennettTypesBasedChannelSimulationQuantumReverseShannonTheorem,2020TransITExactChannelSynthesis,2024ISITSharangAaronOptimalSecondOrderTermForChannelSynthesis},
as depicted in Figure \ref{fig:block_coding_channel_synthesis}.
The problem is then to determine the smallest compression rate $R$ in bits per sample sufficient to achieve
\begin{IEEEeqnarray}{c}
\Delta(P_{X^n,Y^n}, q_{X,Y}^{\otimes n}) \underset{n \to \infty}{\longrightarrow} 0,\label{eq:def_strong_coordination}
\end{IEEEeqnarray}
where
$q_{X,Y}$ is the target joint distribution,
and
$\Delta$ is some divergence.
The answer has been shown to depend on the amount of available common randomness.
The problem in the absence of common randomness was first studied in \cite{1975WynerCommonInformationAndSoftCoveringLemma}, under a slightly different formulation. In that case, the optimal rate is the Wyner common information $C_q(X,Y).$
The case of unlimited common randomness was considered\\

{\small
This research was supported by the United Kingdom Engineering and Physical Sciences Research Council (EPSRC) for the projects AIR (ERC Consolidator Grant, EP/X030806/1) and INFORMED-AI (EP/Y028732/1). For the purpose of open access, the authors have applied a Creative Commons Attribution (CCBY) license to any Author Accepted Manuscript version arising from this submission.
}
\newpage
\noindent
in \cite{2002BennettReverseShannonTheorem,2002WinterChannelSimulationUnlimitedCR},
where the optimal rate $R$ is found to be the Shannon mutual information $I_q(X;Y).$
Bounds on the optimal average coding
length in the one-shot setting (Figure \ref{fig:one_shot_channel_simulation}) have been obtained in \cite{2007HarshaOneShotChannelSimulationNearlySameRateAsPoisson,2018PoissonFunctionalRepresentationLemma}.
The role of common randomness is characterized in \cite{2014BennettWinterChannelSimulationFiniteRateCRWorstCaseSource,2013PaulCuffDistributedChannelSynthesis} --- see also \cite{Cuff2010CoordinationCapacity}.
In \cite{Cuff2010CoordinationCapacity,2013PaulCuffDistributedChannelSynthesis}, the minimum rate of common randomness that is necessary to reach the optimal compression rate $I_q(X;Y)$ was coined \textit{necessary conditional entropy}. Notably, it is often equal to the Shannon conditional entropy $H_q(Y|X),$ and thus often larger than the optimal
compression rate $I_q(X;Y).$ This poses a significant implementation challenge.
Therefore, it is more practically relevant to assume that the rate of common randomness is limited.
See also \cite{2020GregFlamishNeuripsFirstDefinitionRelativeEntropyCoding,2022GregAStarCodingforRelativeEntropyCoding,2023GregOptimalRuntimeChannelSimulationFor1DUnimodal,2023GregGeneralAccelerationRelativeEntropyCodingByGreedyRejection,2023GregTheisAdaptiveGreedyRejectionSampling,2024GregSpacePartitioningToAccelerateRelativeEntropyCoding} for progress on the computational efficiency of algorithms for the one-shot version of the problem (Figure \ref{fig:one_shot_channel_simulation}).



The problem of channel synthesis has been shown to be applicable to a broad range of lossy source coding problems, as an alternative to quantization.
In neural compression, it provides a differentiable alternative to standard quantization which has the merit of
providing more control on the compression noise and of making use of any potential common randomness between the encoder and the decoder
\cite{Havasi2019MinimalRandomCoding,2020GregRelativeEntropyCoding,2020TheisAgustssonDitheredQuantization,2022TheisDiffusionWithChannelSimulation}.
In federated learning, channel synthesis has been shown to allow state-of-the-art communication costs
\cite{2023FrancescoPaseGunduzChannelSimulationForFederatedLearning, 2024FrancescoPaseGunduzChannelSimulationForFederatedLearningAdaptive}.
Furthermore, channel synthesis is amenable to providing differential privacy guarantees
\cite{2021ChristosChannelSimulationInFederatedLearningForDifferentialPrivacy, 2024BurakGunduzDitheredQuantizationForFederatedLearning, 2024DieuleveutChannelSimulationFederatedLearning, 2024CTLiChannelSimulationFederatedLearningLaplace, Liu:NeurIPS:24}. Please refer to \cite{Li:FT:24} for a comprehensive overview of the channel synthesis problem and its applications. 

\begin{figure}[t!]
\centering\includegraphics[width=0.48\textwidth]{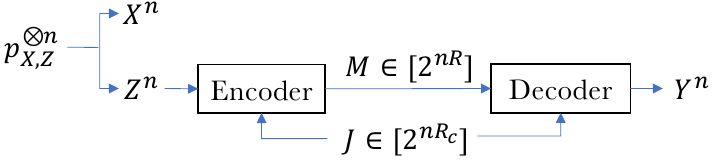}
\caption{Our remote channel synthesis setup.}
\label{fig:our_setup}
\end{figure}

In this paper, we consider the problem of \textit{remote channel synthesis} (r.c.s.), depicted in Figure \ref{fig:our_setup}. It differs from the the standard channel synthesis problem (Figure \ref{fig:block_coding_channel_synthesis}) in that the encoder only has access to a partial or noisy version $Z^n$ of
\newpage
\noindent
the remote sequence $X^n,$ such that $(X^n,Z^n)$ follows an i.i.d. distribution
$q_{X,Z}^{\otimes n}.$
This problem bears a resemblance with the remote rate-distortion problem \cite{1962DobrushinTsybakovFirstRemoteRateDistortion,1970WolfZivRemoteRateDistortion},
\cite[Chap.~3,~Sec.~5]{1971BookBergerRateDistortionTheory},
\cite{1980WitsenhausenIndirectRateDistortion}, 
in which a constraint of the form
\begin{IEEEeqnarray}{c}
\mathbb{E}[d(X^n,Y^n)] \leq \Delta
\nonumber
\end{IEEEeqnarray}
is imposed instead of \eqref{eq:def_strong_coordination},
where $d$ is a measure of distortion.
We obtain a single-letter characterization of the optimal rate pairs $(R,R_c),$ where $R$ denotes the compression rate and $R_c$ the common randomness rate. We find that in the more practically relevant setting with a small $R_c,$ coordinating $Z^n$ and $Y^n$ such that
\begin{IEEEeqnarray}{c}
\Delta(P_{Z^n,Y^n}, q_{Z,Y}^{\otimes n}) \underset{n \to \infty}{\longrightarrow} 0
\end{IEEEeqnarray}
for some $q_{Z,Y}$
is sub-optimal. In other words, in order to achieve the optimal rate $R,$ it is necessary --- in most cases --- to employ a vector scheme, in the sense that the joint distribution of $Z^n$ and $Y^n$ cannot be arbitrarily close to an i.i.d. distribution $q_{Z,Y}^{\otimes n}$ --- or more generally a product distribution.

Our problem is part of the broader literature on \textit{distributed coordination} \cite{1975WynerCommonInformationAndSoftCoveringLemma,1993TITHanVerduApproximationOfOutputStatistics,1973GacsKornerCommonInformation}. This includes problems involving more than two terminals. Communication links between the latter may also be noisy.
Channel synthesis is part of a subset of such problems involving coordination between sources and outputs \cite{2002BennettReverseShannonTheorem,Cuff2010CoordinationCapacity,2011ITWorkshopGohariAnantharamGeneratingDependentRVsOverNetworks,2013PaulCuffDistributedChannelSynthesis,2015TransITYassaeeGohariInteractiveCommunication,2018MaelLeTreustCausalStrongCoordination,2020TransITExactChannelSynthesis,2020WatanabeSurvey,2023PradhanStrongCoordinationOneDecoderMultipleNonInteractingEncoders,2024ISITCheukTingLiOneShotCoordinationOnNetworkWithOneFinalOutput,2024ISITCoordinatingOneSourceWithOutputsOfDecodersCommonMessageSingleEncoder,2024TranITStrongCoordinationOfSingleSourceWithMultipleDecoderOutputs,2024ISITSharangAaronOptimalSecondOrderTermForChannelSynthesis}.
A channel coding problem with a name similar to our problem's has been studied in
\cite{2020CerviaRemoteJointStrongCoordination}.
While that channel coding problem involves a coordination constraint similar to our remote coordination constraint, the problem is fundamentally different: the (channel coding) rate is a function of the target joint distribution,
while our problem consists in characterizing the optimal (compression) rate for a given target joint distribution.

While our work focuses on the strong coordination constraint in \eqref{eq:def_strong_coordination}, other types of coordination constraints have been considered, such as
\begin{IEEEeqnarray}{c}
\Delta(\hat{P}_{X_{1:n}^{(1)}, \ldots, X_{1:n}^{(k)}}, q_{X^{(1)}, \ldots, X^{(k)}}) \overset{\mathcal{P}}{\underset{n \to \infty}{\longrightarrow}} 0,\label{eq:def_empirical_coordination}
\end{IEEEeqnarray}
\begin{IEEEeqnarray}{c}
\Delta(P_{X^{(1)}, \ldots, X^{(k)}}, q_{X^{(1)}, \ldots, X^{(k)}}) \underset{n \to \infty}{\longrightarrow} 0\label{eq:def_asymptotic_coordination_with_single_symbol_outputs}
\end{IEEEeqnarray}
where
$X^{(i)}_{1:n}$ denotes a sequence of length $n,$ and
$\hat{P}$ denotes the joint type \cite{2011BookElGamalKimNetworkInformationTheory}, and $\overset{\mathcal{P}}{\to}$ stands for convergence in probability. The constraint \eqref{eq:def_empirical_coordination} is \textit{empirical coordination constraint}.
Constraints in the form of \eqref{eq:def_asymptotic_coordination_with_single_symbol_outputs} are called \textit{correlation distillation} constraints in the case where $q$ satisfies $X^{(1)} = \cdots = X^{(k)}.$
\newpage
\noindent
Constraints \eqref{eq:def_strong_coordination}, \eqref{eq:def_empirical_coordination}, and \eqref{eq:def_asymptotic_coordination_with_single_symbol_outputs} are instances of \textit{approximate coordination} constraints, as opposed to
\textit{exact coordination}.
Alternatively, the divergence may simply be required to be bounded
\cite{2014ImperfectCoordinationNonInteractiveDesireOftenIdenticalOutputs}. 
The problem formulation is laid out in Section \ref{sec:problem_formulation}.
We present our main results in Section \ref{sec:main_results}. The proofs can be found
in the appendices.
In Section \ref{sec:scalar_vs_vector}, we provide additional remarks on the necessity of vector schemes to reach the optimal compression rate in the case where the rate of common randomness is small.

\section{Problem formulation}\label{sec:problem_formulation}

\subsection{Notation}
Calligraphic letters such as $\mathcal{X}$ denote sets, except in $p^{\mathcal{U}}_{\mathcal{J}},$ which denotes the uniform distribution over alphabet $\mathcal{J}.$ 
We denote by $[a]$ the set $\{1, ..., \lfloor a \rfloor\},$ and by $x^n$ the finite sequence $(x_1, ..., x_n).$ 
We denote by $\|p-q\|_{TV}$ the total variation distance between distributions $p$ and $q.$ The closure of a set is denoted by an over-lined letter, e.g. $\overline{\mathcal{A}}.$ Given finite-valued random variables $X$ and $Y$ with distribution $p_{X,Y}$ we use $H_p(X)$ to denote the Shannon entropy of $X$ and $I_p(X;Y)$ to
denote
\noindent
the Shannon mutual information between $X$ and $Y.$

\subsection{Definitions}

Throughout this paper, the alphabet of the remote source is denoted by $\mathcal{X},$ that of the observed source is denoted by $\mathcal{Z},$ and that of the decoder output is denoted by $\mathcal{Y}.$ We assume said alphabets to be finite.

\begin{definition}\label{def:code_remote_strong_coordination}
Given
a positive integer $n,$
and non-negative reals $R$ and $R_c,$
an $(n, R, R_c)$
code
is a privately randomized encoder and decoder pair $(F^{(n)},G^{(n)})$ consisting of a conditional probability kernel $F^{(n)}_{M|\textcolor{black}{Z^n}, J}$
from $ \textcolor{black}{\mathcal{Z}^n} \times [2^{nR_c}] $ to $ [2^{nR}],$ and a conditional probability kernel $G^{(n)}_{Y^n|J, M}$ from $ [2^{nR_c}] \times [2^{nR}] $ to $\mathcal{Y}^n.$
Given a distribution $q_{X,\textcolor{black}{Z}}$ on $\mathcal{X}\textcolor{black}{\times \mathcal{Z}},$ the following distribution
\begin{IEEEeqnarray}{c}
P_{X^n, \textcolor{black}{Z^n,} J, M, Y^n} := q_{X\textcolor{black}{,Z}}^{\otimes n} \cdot p^{\mathcal{U}}_{[2^{nR_c}]} \cdot F^{(n)} \cdot G^{(n)}\nonumber
\end{IEEEeqnarray}is called \textit{the distribution induced by the code}.
\end{definition}
\begin{definition}\label{def:achievability_remote_strong_coordination}
Given
a distribution $q_{X,Z}$ on $\mathcal{X} \times \mathcal{Z},$
and a distribution $q_{X,Y}$ on $\mathcal{X} \times \mathcal{Y},$
a rate pair $(R, R_c)$ is
$(\textcolor{black}{q_{X,Z},}q_{X,Y})$-achievable if there exists a sequence $\{(F^{(n)},G^{(n)})\}_{n\in\mathbb{N}}$ of codes, the $n$-th being $(n, R, R_c),$ such that 
\begin{IEEEeqnarray}{c}
\|\textcolor{black}{P_{X^n,Y^n} - q_{X,Y}^{\otimes n}}\|_{TV} \underset{n \to \infty}{\longrightarrow} 0.\label{eq:def_achievability_remote_strong_coordination_constraint}
\end{IEEEeqnarray}
$\mathcal{A}^{(\textcolor{black}{r.c.s.})}$ denotes the set of $(\textcolor{black}{q_{X,Z},}q_{X,Y})$-achievable rate pairs $(R,R_c),$ where r.c.s. stands for remote channel synthesis.
\end{definition}

The following lemma is a straightforward reformulation of Definition \ref{def:achievability_remote_strong_coordination}.
\begin{lemma}\label{lemma:conditions_for_P_to_define_a_code}
Consider
a joint distribution $q_{X,Z}$ on $\mathcal{X} \times \mathcal{Z},$
a joint distribution $q_{X,Y}$ on $\mathcal{X} \times \mathcal{Y},$
and
a distribution $P_{X^n, \textcolor{black}{Z^n,} J, M, Y^n}$ 
on $\mathcal{X}^n \textcolor{black}{\times \mathcal{Z}^n} \times \mathbb{N} \times \mathbb{N} \times \mathcal{Y}^n.$ Suppose that there exist non-negative reals $R$ and $R_c$ such that $M \in [2^{nR}]$ and $J \in [2^{nR_c}],$ $P$-almost surely. Then, $P$ coincides with the
distribution induced by the $(n,R,R_c)$ code
$(P_{M|\textcolor{black}{Z^n},J},P_{Y^n|J,M})$
\newpage
\noindent
if and only if $P$ satisfies the following properties:
\begin{itemize}
    \item $P_{X^n\textcolor{black}{,Z^n},J} \equiv q_{X\textcolor{black}{,Z}}^{\otimes n} \cdot p^{\mathcal{U}}_{[2^{nR_c}]}$
    \item $\textcolor{black}{X^n-Z^n}-(J,M)-Y^n$ form a Markov chain
\end{itemize}
\end{lemma}

\section{Main results}\label{sec:main_results}

The main result of this paper is stated in the following theorem.
\begin{theorem}\label{thm:remote_strong_coordination_region}
Consider
a joint distribution $q_{X,Z}$ on $\mathcal{X} \times \mathcal{Z},$
and
a joint target distribution $q_{X,Y}$ on $\mathcal{X} \times \mathcal{Y}.$
Define the region $\mathcal{S}^{(\textcolor{black}{r.c.s.})}$ as 
\begin{align}\label{eq:def_S_remote_strong_coordination}
      & 
    \left\{ \begin{array}{rcl}
        \scalebox{0.9}{$(R, R_c) \in \mathbb{R}_{\geq 0}^2$} &:& \scalebox{0.9}{$\exists \ p_{X,Z,W,Y} \in \mathcal{D}^{(\textcolor{black}{r.c.s.})},$}\\
        \scalebox{0.9}{$R$} &\geq& \scalebox{0.9}{$\textcolor{black}{I_p(Z;W)}$} \\
        \scalebox{0.9}{$R+R_c$} &\geq& \scalebox{0.9}{$\textcolor{black}{I_p(X,Y;W)}$}
    \end{array}\right\},
\end{align} with $\mathcal{D}^{(\textcolor{black}{r.c.s.})}$ defined as
\begin{align}\label{eq:def_D_remote_strong_ccordination}
       & 
    \left\{ \begin{array}{rcl}
        &p_{X,Z,W,Y} : p_{X,Z} \equiv q_{X,Z}, \ \textcolor{black}{p_{X,Y} \equiv q_{X,Y}}&\\
        &\textcolor{black}{X - Z - W - Y}&\\
        &\textcolor{black}{|\mathcal{W}| \leq |\mathcal{Y}||\mathcal{Z}|+1}&
    \end{array}\right\}.
\end{align}Then, the closure of
$\mathcal{A}^{(\textcolor{black}{r.c.s.})}$
is equal to $\mathcal{S}^{(\textcolor{black}{r.c.s.})}.$
\end{theorem}

The proof is very close to that in \cite{2013PaulCuffDistributedChannelSynthesis} for the standard channel synthesis problem, and can be found in
Appendices \ref{app:converse_remote_strong_coordination} and \ref{app:achievability_thm_r_s_c}.
The scheme uses the likelihood encoder \cite{2016CuffLikelihoodEncoder}, which has been shown to be universal in that it is relevant for numerous lossy source compression problems.
In contrast, the traditional scheme for the similar problem of remote rate-distortion is not universal.
Note that $\mathcal{S}^{(r.c.s.)}$ is non-empty if and only if $(q_{X,Z},q_{X,Y})$ belongs to the set $\mathcal{F}^{(r.c.s.)}$ defined as
\begin{IEEEeqnarray}{c}
\{(q_{X,Z},q_{X,Y}) \ | \ \exists q_{Y|Z}, \ q_{X,Y} \equiv \sum_{z\in\mathcal{Z}} q_Z(z)q_{X|Z=z}q_{Y|Z=z}\}.
\nonumber
\end{IEEEeqnarray}
A pair $(q_{X,Z},q_{X,Y})$ is said to be \textit{feasible} if it belongs to $\mathcal{F}^{(r.c.s.)}.$ When the condition in the latter's definition is satisfied by some conditional distribution $q_{Y|Z},$ we say that the resulting distribution $q_{X,Z}\cdot q_{Y|Z}$ is \textit{compatible with} $(q_{X,Z},q_{X,Y}).$
A natural scheme for remote channel synthesis consists in synthesizing such a compatible channel $q_{Y|Z}.$ We call this direct channel synthesis, abbreviated as d.c.s. This leads to the following achievable region.

\begin{proposition}(Direct channel synthesis)\\
Consider a $(q_{X,Z},q_{X,Y})$-achievable rate pair $(R,R_c)$ and a distribution $q_{X,Z,Y}$ compatible with $(q_{X,Z},q_{X,Y}).$
Define the region
$\mathcal{S}^{(d.c.s.)}(q_{X,Z,Y})$ as 
\begin{align}\label{eq:def_S_direct_strong_coordination}
      & 
    \left\{ \begin{array}{rcl}
        \scalebox{0.9}{$(R, R_c) \in \mathbb{R}_{\geq 0}^2$} &:& \scalebox{0.9}{$\exists \ p_{X,Z,W,Y} \in \mathcal{D}^{(d.c.s.)}(q_{X,Z,Y}),$}\\
        \scalebox{0.9}{$R$} &\geq& \scalebox{0.9}{$I_p(Z;W)$} \\
        \scalebox{0.9}{$R+R_c$} &\geq& \scalebox{0.9}{$I_p(Z,Y;W)$}
    \end{array}\right\},
\end{align} with $\mathcal{D}^{(d.c.s.)}(q_{X,Z,Y})$ defined as
\begin{align}\label{eq:def_D_direct_strong_ccordination}
       & 
    \left\{ \begin{array}{rcl}
        &p_{X,Z,W,Y} :
        p_{X,Z,Y} \equiv q_{X,Z,Y}
        &\\
        &X - Z - W - Y&\\
        &|\mathcal{W}| \leq |\mathcal{Y}||\mathcal{Z}|+1&
    \end{array}\right\}.
\end{align}
\newpage
\noindent
Then, $(R,R_c)\in\mathcal{S}^{(d.c.s.)}(q_{X,Z,Y})$ iff there exists $\varepsilon>0$ and a sequence $\{(F^{(n)},G^{(n)})\}_{n\in\mathbb{N}}$ of codes, the $n$-th being
$(n, R+\varepsilon, R_c),$ such that 
\begin{IEEEeqnarray}{c}
\|P_{Z^n,Y^n} - q_{Z,Y}^{\otimes n}\|_{TV} \underset{n \to \infty}{\longrightarrow} 0.\label{eq:in_prop_analog_r_s_c_analogueness}
\end{IEEEeqnarray}
When \eqref{eq:in_prop_analog_r_s_c_analogueness} holds, then, from the Markov chain relation $X-Z-W-Y,$ and \cite[Lemma~V.2]{2013PaulCuffDistributedChannelSynthesis}, we have
\begin{IEEEeqnarray}{c}
\|P_{X^n,Z^n,Y^n} - q_{X,Z,Y}^{\otimes n}\|_{TV} \underset{n \to \infty}{\longrightarrow} 0.\label{eq:in_prop_analog_r_s_c_correctness}
\end{IEEEeqnarray}
Hence, we always have $\mathcal{S}^{(d.c.s.)}(q_{X,Z,Y}) \subseteq \overline{\mathcal{A}}^{(r.c.s.)} = \mathcal{S}^{(r.c.s.)}.$
\end{proposition}
\begin{IEEEproof}
The single-letter characterization follows from the single-letter characterization \cite[Theorem~II.1]{2013PaulCuffDistributedChannelSynthesis} for the problem of channel synthesis.
Indeed, the proof of the latter involves an arbitrarily small excess rate $\varepsilon$ for the compression rate, but not for the common randomness rate.
The convergence in \eqref{eq:in_prop_analog_r_s_c_correctness} follows from the Markov chain relation $X-Z-W-Y$ and \cite[Lemma~V.2]{2013PaulCuffDistributedChannelSynthesis}.
The equality $\overline{\mathcal{A}}^{(r.c.s.)} = \mathcal{S}^{(r.c.s.)}$ follows from Theorem \ref{thm:remote_strong_coordination_region}.
\end{IEEEproof}

\begin{figure}[t!]
\centering\includegraphics[width=0.48\textwidth]{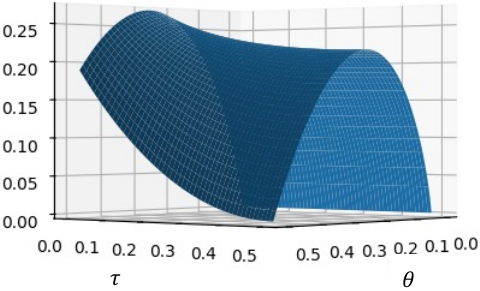}
\caption{
Lower bound on the gap between the optimal rates in \eqref{eq:statement_suboptimality_of_d_s_c} --- see
Appendix \ref{app:proof_binary_strict_suboptimality_of_direct_strong_coordination}
for details.
Variables $\theta$ and $\tau$ denote $q_{Z,Y}(Y\neq Z)$ and $q_{X,Z}(X\neq Z),$ respectively, where $q_{X,Z,Y}$ is a distribution compatible with $(q_{X,Z},q_{X,Y}).$
}
\label{fig:plot_gap_optimal_rate_remote_coordination_vs_analog}
\end{figure}

For any distribution $p_{X,Z,W,Y}$ satisfying the Markov chain relation $X-Z-W-Y,$ we have
\begin{IEEEeqnarray}{rCl}
I_p(Z,Y;W) &=& I_p(X,Z,Y;W)
\nonumber\\*
&\geq& \max(I_p(Z;W),I_p(X,Y;W)).
\end{IEEEeqnarray}
Therefore, in most non-trivial cases, for small enough common randomness rate $R_c,$ direct channel synthesis is strictly sub-optimal. We show this by obtaining a lower bound on the sub-optimality gap when $R_c=0,$ depicted in Figure \ref{fig:plot_gap_optimal_rate_remote_coordination_vs_analog}.

\begin{proposition}\label{prop:suboptimality_of_of_direct_strong_coordination}
Assume that $(q_{X,Z},q_{X,Y})$ is feasible and
\begin{IEEEeqnarray}{c}
q_X\equiv q_Z \equiv q_Y \equiv Ber(1/2),\nonumber\\*
0<q_{X,Z}(X\neq Z)<q_{X,Y}(X\neq Y)< 1/2.
\end{IEEEeqnarray}
Then, there exists $r>0$ such that for any $R_c\in[0,r),$
\newpage
\begin{IEEEeqnarray}{c}
\min \{R \ | \ (R,R_c) \in \cup_{q_{X,Z,Y}} \mathcal{S}^{(d.c.s.)}(q_{X,Z,Y}) \}
\nonumber\\*
>\min \{R \ | \ (R,R_c) \in \mathcal{S}^{(r.c.s.)}\},\label{eq:statement_suboptimality_of_d_s_c}
\end{IEEEeqnarray}
\noindent
where the union is over all distributions $q_{X,Z,Y}$ on $\mathcal{X}\times\mathcal{Z}\times\mathcal{Y}$ compatible with $(q_{X,Z},q_{X,Y}).$
\end{proposition}

The proof is provided in
Appendix \ref{app:proof_binary_strict_suboptimality_of_direct_strong_coordination}.




\section{Scalar schemes v.s. vector schemes}\label{sec:scalar_vs_vector}

In order to provide further intuitive understanding of Proposition \ref{prop:suboptimality_of_of_direct_strong_coordination},
we consider a slightly different perspective, centered around the concept of \textit{proper vector scheme}, defined below.

\begin{definition}
Consider a source distribution $q_{X,Z},$ a target distribution $q_{X,Y},$ and non-negative reals $R$ and $R_c.$
A \textit{$(R,R_c)$ scheme for distribution $q_{X,Y}$} is a sequence $(F^{(n)},G^{(n)})$ of codes, the $n$-th being $(n,R,R_c),$ such that the corresponding sequence $\{P_{X^n,Z^n,J,M,Y^n}\}_{n\in\mathbb{N}}$ of induced distributions satisfies
\begin{IEEEeqnarray}{c}
\|\textcolor{black}{P_{X^n,Y^n} - q_{X,Y}^{\otimes n}}\|_{TV} \underset{n \to \infty}{\longrightarrow} 0.\label{eq:in_def_notion_of_achievable_scheme_at_rate_R_remote_strong_coordination_constraint}
\end{IEEEeqnarray}
\end{definition}

\begin{definition}
Consider two non-negative reals $R$ and $R_c.$
A $(R,R_c)$ scheme for distribution $q_{X,Y}$ is said to be a \textit{proper vector scheme} if the corresponding sequence $\{P_{X^n,Z^n,J,M,Y^n}\}_{n\in\mathbb{N}}$ of induced distributions is such that there does not exists a sequence $\{\rho^{(t)}_{Y|Z}\}_{t\in\mathbb{N}}$ of single-letter distributions for which
\begin{IEEEeqnarray}{c}
\Big\| P_{Z^n,Y^n} - \prod_{t{=}1}^n q_{Z}\cdot \rho^{(t)}_{Y|Z}\Big\|_{TV} \underset{n \to \infty}{\longrightarrow} 0.
\end{IEEEeqnarray}
\end{definition}




From the results of Section \ref{sec:main_results},
a proper vector scheme is necessary for optimal 
\begin{proposition}
Assume the setting of Proposition \ref{prop:suboptimality_of_of_direct_strong_coordination}.
Let $R_{\min}$ denote the right hand side of \eqref{eq:statement_suboptimality_of_d_s_c} in the case $R_c=0,$ i.e., the optimal rate for remote strong coordination in the absence of common randomness.
Let $R_1$ denote the left hand side of \eqref{eq:statement_suboptimality_of_d_s_c} in the case $R_c=0,$ i.e., the optimal rate for direct strong coordination in the absence of common randomness. Then, for any $R$ in $[R_{\min}, R_1),$ for small enough $R_c,$ any $(R,R_c)$ scheme for distribution $q_{X,Y}$ is a proper vector scheme.
\end{proposition}
\begin{IEEEproof}
Assume the setting of Proposition \ref{prop:suboptimality_of_of_direct_strong_coordination}.
Fix $R$ in $[R_{\min},R_1),$
and $R_c\in[0,R_1-R)$
such that $(R,R_c)$ is $(q_{X,Z},q_{X,Y})$-achievable.
For the sake of contradiction, assume that
there exists
a $(R,R_c)$ scheme for $q_{X,Y}$
which
is not a proper vector scheme.
This scheme can be turned into a $(R+R_c,0)$ scheme for $q_{X,Y}$ having the same induced $P_{Z^n,Y^n},$ by sampling variable $J$ locally at the encoder, and taking $(M,J)$ as message.
In the remainder of the proof, we only study this new scheme.
By assumption and by construction, the latter is not a proper vector scheme.
Then, there exists
a sequence $\{\rho^{(t)}_{Y|Z}\}_{t\in\mathbb{N}}$ of single-letter distributions for which
\newpage
\begin{IEEEeqnarray}{c}
\Big\| P_{Z^n,Y^n} - \prod_{t{=}1}^n q_{Z}\cdot \rho^{(t)}_{Y|Z}\Big\|_{TV} \underset{n \to \infty}{\longrightarrow} 0.\label{eq:in_proof_proper_vector_scheme_assumption_scalar_distributions}
\end{IEEEeqnarray}
Since there are only a finite number of possible single-letter conditional distributions, then (e.g., from \cite[Lemma~V.1]{2013PaulCuffDistributedChannelSynthesis}) $P_{Z^n,Y^n}$ is nearly i.i.d. \textit{by part}. We show that there exits a $(R+R_c,0)$ scheme for $q_{X,Y}$ which performs direct strong coordination by synthesizing one such conditional distribution.
First, we have
\begin{claim}
For any $t\in\mathbb{N},$ the conditional distribution $\rho^{(t)}_{Y|Z}$ is compatible with $(q_{X,Z},q_{X,Y}).$
\end{claim}
\begin{IEEEproof}
From \eqref{eq:in_proof_proper_vector_scheme_assumption_scalar_distributions}, from \cite[Lemma~V.2]{2013PaulCuffDistributedChannelSynthesis}, we have
\begin{IEEEeqnarray}{c}
\Big\| P_{X^n,Z^n,Y^n} - \prod_{t{=}1}^n q_{X,Z}\cdot \rho^{(t)}_{Y|Z}\Big\|_{TV} \underset{n \to \infty}{\longrightarrow} 0.
\end{IEEEeqnarray}
Thus, from \cite[Lemma~V.1]{2013PaulCuffDistributedChannelSynthesis}, we have
\begin{IEEEeqnarray}{c}
\Big\| P_{X^n,Y^n}
- \prod_{t{=}1}^n \Big[
\sum_{z} q_Z(z)q_{X|Z{=}z}\cdot \rho^{(t)}_{Y|Z{=}z}
\Big]
\Big\|_{TV} \underset{n \to \infty}{\longrightarrow} 0.
\nonumber
\end{IEEEeqnarray}
Since the scheme is an $(R,R_c)$ scheme for $q_{X,Y},$
then,
from the triangle inequality for the total variation distance, we have
\begin{IEEEeqnarray}{c}
\Big\| q_{X,Y}^{\otimes n}
- \prod_{t{=}1}^n \Big[
\sum_{z} q_Z(z)q_{X|Z{=}z}\cdot \rho^{(t)}_{Y|Z{=}z}
\Big]
\Big\|_{TV} \underset{n \to \infty}{\longrightarrow} 0.
\nonumber
\end{IEEEeqnarray}
For any $t\in\mathbb{N},$ from
\cite[Lemma~V.1]{2013PaulCuffDistributedChannelSynthesis}
with $U=(X_t,Y_t)$ therein,
we obtain
\begin{IEEEeqnarray}{c}
\Big\| q_{X,Y}
-
\sum_{z} q_Z(z)q_{X|Z{=}z}\cdot \rho^{(t)}_{Y|Z{=}z}
\Big\|_{TV} \underset{n \to \infty}{\longrightarrow} 0.
\nonumber
\end{IEEEeqnarray}
Since this quantity does not depend on $n,$ it is null. Hence, $\rho^{(t)}_{Y|Z}$ is compatible with $(q_{X,Z},q_{X,Y}).$
\end{IEEEproof}
\vspace{5pt}

Second, we have
\begin{claim}\label{claim:crux_of_proof_proper_vector_schemes}
There exits $t\in\mathbb{N}$ such that
\begin{IEEEeqnarray}{c}
(R+R_c,0)\in\mathcal{S}^{(c.s.)}(\rho^{(t)}_{Y|Z}),
\end{IEEEeqnarray}
where, for a conditional distribution $\mu_{Y|Z},$ the set $\mathcal{S}^{(c.s.)}(\mu_{Y|Z})$ is defined as the achievable region for the standard channel synthesis problem \cite[Eq~8~\&~9]{2013PaulCuffDistributedChannelSynthesis}
for channel $\mu_{Y|Z}$ and source distribution $q_Z.$
\end{claim}
\begin{IEEEproof}
We present a sketch of the proof.
The technical details are
very similar to those of the converse proof of Theorem \ref{thm:remote_strong_coordination_region}.
Following steps similar to those in the converse proof of \cite[Section~VI]{2013PaulCuffDistributedChannelSynthesis} for the standard problem of channel synthesis, we prove the following.
For every $k,$ there exits $t_k$ and $n_k$ such that
\begin{IEEEeqnarray}{c}
\|P_{Z_T,Y_T} - q_Z \cdot \rho^{(t_k)}_{Y|Z}\|_{TV} \leq 1/k
\end{IEEEeqnarray}
and
\begin{IEEEeqnarray}{c}
R+R_c \geq I_{P}(W;Z_T,Y_T) - g(1/k),
\end{IEEEeqnarray}
\newpage
\noindent
where $P$ denotes the distribution induced by the $n_k$-th code of the scheme, $T$ is a certain random variable taking values in $[n_k],$ and $g:\mathbb{R}_+ \to \mathbb{R}$ is a function with limit $0$ at $0.$
From a continuity argument similar to that in \cite[Section~VI]{2013PaulCuffDistributedChannelSynthesis}, and from the fact that there exist only a finite number of different conditional distributions $\rho_{Y|Z},$ we obtain the desired conclusion.
\end{IEEEproof}

The above two claims imply that $(R+R_c,0)\in\mathcal{S}^{(d.c.s.)}(q_{X,Z}\cdot \rho^{(t)}_{Y|Z})$ for a certain $t\in\mathbb{N}.$
From \eqref{eq:statement_suboptimality_of_d_s_c},
this contradicts the assumption that $R+R_c < R_1,$
which concludes the proof.
\end{IEEEproof}

\section{Conclusion}

We have considered the problem of remote channel synthesis, where an encoder has access to a partial or noisy version $Z^n$ of a remote source $X^n,$ and communicates information to a decoder which aims to produce an output $Y^n$ coordinated with the remote source in the strong sense.
We have obtained a single-letter characterization of the optimal compression and common randomness rate pairs. Owing to practical implementation challenges, the rate of common randomness if often relatively small. We find that
when it is small enough,
then in most cases,
the natural approach consisting in
achieving a joint distribution of $Z^n$ and $Y^n$ that approaches a product distribution asymptotically is strictly sub-optimal.
We expect that the fundamental problem of remote channel synthesis can be used as a building block for more complex distributed coordination schemes.




\IEEEtriggeratref{25}

\newpage \hfill
\newpage
\bibliographystyle{IEEEtran}
\bibliography{biblio}

\begin{thebibliography}{10}
\providecommand{\url}[1]{#1}
\csname url@samestyle\endcsname
\providecommand{\newblock}{\relax}
\providecommand{\bibinfo}[2]{#2}
\providecommand{\BIBentrySTDinterwordspacing}{\spaceskip=0pt\relax}
\providecommand{\BIBentryALTinterwordstretchfactor}{4}
\providecommand{\BIBentryALTinterwordspacing}{\spaceskip=\fontdimen2\font plus
\BIBentryALTinterwordstretchfactor\fontdimen3\font minus
  \fontdimen4\font\relax}
\providecommand{\BIBforeignlanguage}[2]{{%
\expandafter\ifx\csname l@#1\endcsname\relax
\typeout{** WARNING: IEEEtran.bst: No hyphenation pattern has been}%
\typeout{** loaded for the language `#1'. Using the pattern for}%
\typeout{** the default language instead.}%
\else
\language=\csname l@#1\endcsname
\fi
#2}}
\providecommand{\BIBdecl}{\relax}
\BIBdecl

\bibitem{2020GregFlamishNeuripsFirstDefinitionRelativeEntropyCoding}
G.~Flamich \emph{et~al.}, ``{Compressing Images by Encoding Their Latent
  Representations with Relative Entropy Coding},'' in \emph{{Annual Conference
  on Neural Information Processing Systems}}, 2020, Conference Paper.

\bibitem{2002BennettReverseShannonTheorem}
C.~H. Bennett \emph{et~al.}, ``{Entanglement-assisted capacity of a quantum
  channel and the reverse Shannon theorem},'' \emph{IEEE Transactions on
  Information Theory}, vol.~48, no.~10, pp. 2637--2655, 2002.

\bibitem{2013PaulCuffDistributedChannelSynthesis}
P.~Cuff, ``{Distributed Channel Synthesis},'' \emph{{IEEE Transactions on
  Information Theory}}, vol.~59, no.~11, 2013.

\bibitem{2014BennettTypesBasedChannelSimulationQuantumReverseShannonTheorem}
C.~H. Bennett \emph{et~al.}, ``{The Quantum Reverse Shannon Theorem and
  Resource Tradeoffs for Simulating Quantum Channels},'' \emph{IEEE
  Transactions on Information Theory}, vol.~60, no.~5, pp. 2926--2959, 2014.

\bibitem{2020TransITExactChannelSynthesis}
L.~Yu and V.~Y.~F. Tan, ``{Exact Channel Synthesis},'' \emph{IEEE Transactions
  on Information Theory}, vol.~66, no.~5, pp. 2799--2818, 2020.

\bibitem{2024ISITSharangAaronOptimalSecondOrderTermForChannelSynthesis}
S.~M. Sriramu and A.~B. Wagner, ``{Optimal Redundancy in Exact Channel
  Synthesis},'' in \emph{IEEE International Symposium on Information Theory
  (ISIT)}, 2024.

\bibitem{1975WynerCommonInformationAndSoftCoveringLemma}
A.~Wyner, ``The common information of two dependent random variables,''
  \emph{IEEE Transactions on Information Theory}, vol.~21, no.~2, pp. 163--179,
  1975.

\bibitem{2002WinterChannelSimulationUnlimitedCR}
A.~Winter, ``{Compression of Sources of Probability Distributions and Density
  Operators},'' 2002, arXiv:quant-ph/0208131.

\bibitem{2007HarshaOneShotChannelSimulationNearlySameRateAsPoisson}
P.~Harsha \emph{et~al.}, ``{The Communication Complexity of Correlation},'' in
  \emph{{IEEE Conference on Computational Complexity}}, 2007, Conference
  Proceedings.

\bibitem{2018PoissonFunctionalRepresentationLemma}
C.~T. Li and A.~E. Gamal, ``{Strong Functional Representation Lemma and
  Applications to Coding Theorems},'' \emph{IEEE Transactions on Information
  Theory}, vol.~64, no.~11, 2018.

\bibitem{2014BennettWinterChannelSimulationFiniteRateCRWorstCaseSource}
C.~H. Bennett \emph{et~al.}, ``{The Quantum Reverse Shannon Theorem and
  Resource Tradeoffs for Simulating Quantum Channels},'' \emph{IEEE
  Transactions on Information Theory}, vol.~60, no.~5, 2014.

\bibitem{Cuff2010CoordinationCapacity}
P.~W. Cuff \emph{et~al.}, ``{Coordination capacity},'' \emph{IEEE Transactions
  on Information Theory}, vol.~56, no.~9, pp. 4181--4206, 2010.

\bibitem{2022GregAStarCodingforRelativeEntropyCoding}
G.~Flamich, S.~Markou, and J.~M. Hernandez-Lobato, ``{Fast Relative Entropy
  Coding with A* Coding},'' in \emph{International Conference on Machine
  Learning}, 2022, Conference Paper.

\bibitem{2023GregOptimalRuntimeChannelSimulationFor1DUnimodal}
G.~Flamich, ``{Greedy Poisson Rejection Sampling},'' in \emph{{Annual
  Conference on Neural Information Processing Systems}}, 2023, Conference
  Paper.

\bibitem{2023GregGeneralAccelerationRelativeEntropyCodingByGreedyRejection}
G.~Flamich, S.~Markou, and J.~M. Hernández-Lobato, ``{Faster Relative Entropy
  Coding with Greedy Rejection Coding},'' in \emph{{Annual Conference on Neural
  Information Processing Systems}}, 2023, Conference Paper.

\bibitem{2023GregTheisAdaptiveGreedyRejectionSampling}
G.~Flamich and L.~Theis, ``{Adaptive Greedy Rejection Sampling},'' in
  \emph{IEEE International Symposium on Information Theory}, 2023, Conference
  Paper.

\bibitem{2024GregSpacePartitioningToAccelerateRelativeEntropyCoding}
H.~Jiajun, F.~Gergely, and H.-L. Jos~Miguel, ``{Accelerating Relative Entropy
  Coding with Space Partitioning},'' 2024, Conference Paper.

\bibitem{Havasi2019MinimalRandomCoding}
H.~Marton \emph{et~al.}, ``{Minimal Random Code Learning: Getting Bits Back
  from Compressed Model Parameters},'' in \emph{International Conference on
  Learning Representations}, 2019.

\bibitem{2020GregRelativeEntropyCoding}
G.~Flamich \emph{et~al.}, ``{Compressing Images by Encoding Their Latent
  Representations with Relative Entropy Coding},'' in \emph{{Annual Conference
  on Neural Information Processing Systems}}, 2020.

\bibitem{2020TheisAgustssonDitheredQuantization}
E.~Agustsson and L.~Theis, ``{Universally Quantized Neural Compression},'' in
  \emph{{Annual Conference on Neural Information Processing Systems}}, 2020.

\bibitem{2022TheisDiffusionWithChannelSimulation}
L.~Theis \emph{et~al.}, ``{Lossy Compression with Gaussian Diffusion},'' 2022,
  arXiv:2206.08889.

\bibitem{2023FrancescoPaseGunduzChannelSimulationForFederatedLearning}
B.~Isik \emph{et~al.}, ``{Sparse Random Networks for Communication-Efficient
  Federated Learning},'' in \emph{{International Conference on Learning
  Representations}}, 2023.

\bibitem{2024FrancescoPaseGunduzChannelSimulationForFederatedLearningAdaptive}
------, ``{Adaptive Compression in Federated Learning via Side Information},''
  in \emph{{International Conference on Artificial Intelligence and
  Statistics}}, 2024, Conference Paper.

\bibitem{2021ChristosChannelSimulationInFederatedLearningForDifferentialPrivacy}
A.~Triastcyn, M.~Reisser, and C.~Louizos, ``{Dp-rec: Private \&
  communication-efficient federated learning},'' 2021, arXiv:2111.05454.

\bibitem{2024BurakGunduzDitheredQuantizationForFederatedLearning}
B.~Hasırcıoğlu and D.~Gündüz, ``{Communication efficient private federated
  learning using dithering},'' in \emph{{IEEE International Conference on
  Acoustics, Speech and Signal Processing}}, 2024.

\bibitem{2024DieuleveutChannelSimulationFederatedLearning}
Hegazy \emph{et~al.}, ``{Compression with Exact Error Distribution for
  Federated Learning},'' in \emph{{International Conference on Artificial
  Intelligence and Statistics}}, 2024, Conference Paper.

\bibitem{2024CTLiChannelSimulationFederatedLearningLaplace}
A.~M. Shahmiri, C.~W. Ling, and C.~T. Li, ``{Communication-Efficient Laplace
  Mechanism for Differential Privacy via Random Quantization},'' in \emph{{IEEE
  International Conference on Acoustics, Speech and Signal Processing}}, 2024.

\bibitem{Liu:NeurIPS:24}
Y.~Liu, W.-N. Chen, A.~\"{O}zg\"{u}r, and C.~T. Li, ``Universal exact
  compression of differentially private mechanisms,'' in \emph{Advances in
  Neural Information Processing Systems}, vol.~37, 2024, pp. 91\,492--91\,531.

\bibitem{Li:FT:24}
\BIBentryALTinterwordspacing
C.~T. Li, ``Channel simulation: Theory and applications to lossy compression
  and differential privacy,'' \emph{Foundations and Trends® in Communications
  and Information Theory}, vol.~21, no.~6, pp. 847--1106, 2024. [Online].
  Available: \url{http://dx.doi.org/10.1561/0100000141}
\BIBentrySTDinterwordspacing

\bibitem{1962DobrushinTsybakovFirstRemoteRateDistortion}
R.~Dobrushin and B.~Tsybakov, ``{Information Transmission With Additional
  Noise},'' \emph{IRE Transactions on Information Theory}, vol.~8, no.~5, 1962.

\bibitem{1970WolfZivRemoteRateDistortion}
J.~Wolf and J.~Ziv, ``{Transmission of Noisy Information to a Noisy Receiver
  With Minimum Distortion},'' \emph{IEEE Transactions on Information Theory},
  vol.~16, no.~4, 1970.

\bibitem{1971BookBergerRateDistortionTheory}
T.~Berger, \emph{{Rate Distortion Theory: A Mathematical Basis for Data
  Compression}}.\hskip 1em plus 0.5em minus 0.4em\relax NJ: Prentice Hall:
  Englewood Cliffs, 1971.

\bibitem{1980WitsenhausenIndirectRateDistortion}
H.~Witsenhausen, ``{Indirect Rate Distortion Problems},'' \emph{IEEE
  Transactions on Information Theory}, vol.~26, no.~5, 1980.

\bibitem{1993TITHanVerduApproximationOfOutputStatistics}
T.~Han and S.~Verdu, ``Approximation theory of output statistics,'' \emph{IEEE
  Transactions on Information Theory}, vol.~39, no.~3, pp. 752--772, 1993.

\bibitem{1973GacsKornerCommonInformation}
P.~G{\'a}cs and J.~Korner, ``Common information is far less than mutual
  information.'' \emph{Problems of Control and Information Theory}, vol.~2, pp.
  149--162, 1973.

\bibitem{2011ITWorkshopGohariAnantharamGeneratingDependentRVsOverNetworks}
A.~A. Gohari and V.~Anantharam, ``{Generating dependent random variables over
  networks},'' in \emph{2011 IEEE Information Theory Workshop}, Conference
  Proceedings, pp. 698--702.

\bibitem{2015TransITYassaeeGohariInteractiveCommunication}
M.~H. Yassaee \emph{et~al.}, ``{Channel Simulation via Interactive
  Communications},'' \emph{IEEE Transactions on Information Theory}, vol.~61,
  no.~6, pp. 2964--2982, 2015.

\bibitem{2018MaelLeTreustCausalStrongCoordination}
G.~Cervia \emph{et~al.}, ``{Strong Coordination over Noisy Channels with
  Strictly Causal Encoding},'' in \emph{Annual Allerton Conference on
  Communication, Control, and Computing (Allerton)}, 2018.

\bibitem{2020WatanabeSurvey}
M.~Sudan \emph{et~al.}, ``{Communication for Generating Correlation: A Unifying
  Survey},'' \emph{IEEE Transactions on Information Theory}, vol.~66, no.~1,
  pp. 5--37, 2020.

\bibitem{2023PradhanStrongCoordinationOneDecoderMultipleNonInteractingEncoders}
T.~A. Atif \emph{et~al.}, ``Source coding for synthesizing correlated
  randomness,'' \emph{IEEE Transactions on Information Theory}, vol.~69, no.~1,
  pp. 626--649, 2023.

\bibitem{2024ISITCheukTingLiOneShotCoordinationOnNetworkWithOneFinalOutput}
Y.~Liu and C.~T. Li, ``{One-Shot Coding over General Noisy Networks},'' in
  \emph{IEEE International Symposium on Information Theory (ISIT)}, 2024.

\bibitem{2024ISITCoordinatingOneSourceWithOutputsOfDecodersCommonMessageSingleEncoder}
M.~A. Managoli and V.~M. Prabhakaran, ``{Broadcast Channel Synthesis from
  Shared Randomness},'' in \emph{2024 IEEE International Symposium on
  Information Theory (ISIT)}, Conference Proceedings, pp. 1919--1924.

\bibitem{2024TranITStrongCoordinationOfSingleSourceWithMultipleDecoderOutputs}
M.~X. Cao \emph{et~al.}, ``{Channel Simulation: Finite Blocklengths and
  Broadcast Channels},'' \emph{IEEE Transactions on Information Theory},
  vol.~70, no.~10, pp. 6780--6808, 2024.

\bibitem{2020CerviaRemoteJointStrongCoordination}
G.~Cervia, T.~J. Oechtering, and M.~Skoglund, ``Remote joint strong
  coordination and reliable communication,'' in \emph{2020 IEEE International
  Symposium on Information Theory (ISIT)}, 2020, pp. 932--937.

\bibitem{2011BookElGamalKimNetworkInformationTheory}
A.~El~Gamal and Y.-H. Kim, \emph{{Network Information Theory}}.\hskip 1em plus
  0.5em minus 0.4em\relax {Cambridge (England)}: {Cambridge University Press},
  2011.

\bibitem{2014ImperfectCoordinationNonInteractiveDesireOftenIdenticalOutputs}
S.~O. Chan \emph{et~al.}, ``{On Extracting Common Random Bits From Correlated
  Sources on Large Alphabets},'' \emph{IEEE Transactions on Information
  Theory}, vol.~60, no.~3, pp. 1630--1637, 2014.

\bibitem{2016CuffLikelihoodEncoder}
E.~C. Song, P.~Cuff, and H.~V. Poor, ``The likelihood encoder for lossy
  compression,'' \emph{IEEE Transactions on Information Theory}, vol.~62,
  no.~4, pp. 1836--1849, 2016.

\bibitem{2023YassineGunduzISITRDPSideInformation}
Y.~Hamdi and D.~G\"{u}nd\"{u}z, ``{The Rate-Distortion-Perception Trade-off
  with Side Information},'' in \emph{{IEEE International Symposium on
  Information Theory}}, 2023.

\end{thebibliography}

\newpage
\appendices

\section{Useful lemmas}

\begin{lemma}\label{lemma:TV_joint_to_TV_marginal}\cite[Lemma~V.1]{2013PaulCuffDistributedChannelSynthesis}
    Let $\Pi$ and $\Gamma$ be two distributions on an alphabet $\mathcal{U} \times \mathcal{L}.$ Then \begin{equation*}
        \| \Pi_U - \Gamma_U \|_{TV} \leq \| \Pi_{U,L} - \Gamma_{U,L} \|_{TV}.
    \end{equation*}
\end{lemma}
\begin{lemma}\label{lemma:TV_same_channel}\cite[Lemma~V.2]{2013PaulCuffDistributedChannelSynthesis}
    Let $\Pi$ and $\Gamma$ be two distributions on an alphabet $\mathcal{U} \times \mathcal{L}.$ Then when using the same channel $\Pi_{L|U}$ we have \begin{equation*}
        \| \Pi_U \Pi_{L|U} - \Gamma_U \Pi_{L|U} \|_{TV} = \| \Pi_U - \Gamma_U \|_{TV}.
    \end{equation*}
\end{lemma}
\begin{lemma}\label{lemma:get_expectation_out_of_TV}
    Let $\Pi$ be two distributions on the product of two finite alphabets $\mathcal{U} $ and $\mathcal{L},$ and let $\Pi_{L|U}, \Gamma_{L|U}$ be two channels. Then, we have \begin{equation*}
        \| \Pi_U \Pi_{L|U} - \Pi_U \Gamma_{L|U} \|_{TV} = \mathbb{E}_{\Pi_U} \big[ \| \Pi_{L|U} - \Gamma_{L|U} \|_{TV} \big].
    \end{equation*}
\end{lemma}

\section{Converse direction of Theorem \ref{thm:remote_strong_coordination_region}}\label{app:converse_remote_strong_coordination}

This proof closely tracks \cite[Section~VI]{2013PaulCuffDistributedChannelSynthesis}.\\

\begin{claim}\label{claim:S_is_closed}
$\mathcal{S}^{(\textcolor{black}{r.c.s.})}$ is closed in $\mathbb{R}^2.$
\end{claim}
\begin{IEEEproof}
Note that
$\mathcal{S}^{(\textcolor{black}{r.c.s.})}$ remains unchanged if
variable $W$ in
$\mathcal{D}^{(\textcolor{black}{r.c.s.})}$
is required to take values only in alphabet $\mathcal{W} = [|\mathcal{Y}||\mathcal{Z}|+1].$
Hereafter, we assume that this requirement is added to the definition of $\mathcal{D}^{(\textcolor{black}{r.c.s.})}.$
Then,
$\mathcal{D}^{(\textcolor{black}{r.c.s.})}$
is
a bounded closed subset of a finite-dimensional real vector space. Thus, it
is
compact.
Since the inequalities in the definition of $\mathcal{S}^{(\textcolor{black}{r.c.s.})}$ are not strict inequalities, and mutual information is continuous, and $\mathcal{D}^{(\textcolor{black}{r.c.s.})}$ is closed and bounded,
then for any $(R,R_c) \in \mathcal{S}^{(\textcolor{black}{r.c.s.})},$ the set $\mathcal{D}^{(\textcolor{black}{r.c.s.})}(R,R_c)$ of distributions $p \in \mathcal{D}^{(\textcolor{black}{r.c.s.})}$ such that $R\geq I_p(Z;W)$ and $R+R_c\geq I_p(X,Y;W)$ is a closed and bounded subset of a finite-dimensional real vector space. Therefore, there exists an element $p^{(R,R_c)}$ of that set which is minimal in lexicographic order --- so as not to have to use the axiom of choice later on.
Consider a sequence $\{(R^{(n)},R^{(n)}_c)\}_{n\in\mathbb{N}}$ of elements of $\mathcal{S}^{(\textcolor{black}{r.c.s.})}$ which converges in $\mathbb{R}^2.$ Let $(R^{\infty},R^{\infty}_c)$ be the limit.
From the compactness of $\mathcal{D}^{(\textcolor{black}{r.c.s.})},$ there exists $p^*\in\mathcal{D}^{(\textcolor{black}{r.c.s.})}$ and an increasing sequence $\{n_k\}_{k\in\mathbb{N}}$ of positive integers such that $p^{(R^{(n_k)},R^{(n_k)}_c)} \substack{\raisebox{-4pt}{$\to$} \\ \scalebox{0.5}{$k$$\to$$\infty$}} \; p^*.$
Then, by continuity of the mutual information for finite alphabets,
we have $R^{\infty}\geq I_{p^*}(Z;W)$ and $R^{\infty}+R^{\infty}_c \geq I_{p^*}(X,Y;W).$
Thus, $(R^{\infty},R^{\infty}_c) \in \mathcal{S}^{(\textcolor{black}{r.c.s.})}.$
Hence, $\mathcal{S}^{(\textcolor{black}{r.c.s.})}$ is closed.
\end{IEEEproof}
\vspace{5pt}

We prove that $\overline{\mathcal{A}}^{(\textcolor{black}{r.c.s.})} \subset \mathcal{S}^{(\textcolor{black}{r.c.s.})}$ by proving that $\mathcal{A}^{(\textcolor{black}{r.c.s.})} \subset \mathcal{S}^{(\textcolor{black}{r.c.s.})}.$
Define the region $\mathcal{S}_\varepsilon$ as 
\begin{align}\label{eq:def_S_epsilon}
      & 
    \left\{ \begin{array}{rcl}
        \scalebox{0.9}{$(R, R_c) \in \mathbb{R}_{\geq 0}^2$} &:& \scalebox{0.9}{$\exists \ p_{X,Z,W,Y} \in \mathcal{D}_\varepsilon,$}\\
        \scalebox{0.9}{$R$} &\geq& \scalebox{0.9}{$\textcolor{black}{I_p(Z;W)}$} \\
        \scalebox{0.9}{$R+R_c$} &\geq& \scalebox{0.9}{$\textcolor{black}{I_p(X,Y;W)-2g(\varepsilon)}$}
    \end{array}\right\},
\end{align} with $\mathcal{D}_\varepsilon$ defined as
\begin{align}\label{eq:def_D_epsilon}
       & 
    \left\{ \begin{array}{rcl}
        &p_{X,Z,W,Y} : \textcolor{black}{p_{X,Z} \equiv q_{X,Z},}&\\
        &\|p_{X,Y} - q_{X,Y}\|_{TV} \leq \varepsilon\\
        &\textcolor{black}{X - Z} - W - Y&\\
        &|\mathcal{W}| \leq |\mathcal{Y}|\textcolor{black}{|\mathcal{Z}|}+1&
    \end{array}\right\},
\end{align}
and $g:(0,1/2) \to \mathbb{R}$ defined as
\begin{IEEEeqnarray}{c}
g(\varepsilon) = 4\varepsilon\Big(\log|\mathcal{X}| + \log|\mathcal{Y}| + \log\dfrac{1}{\varepsilon}\Big).
\end{IEEEeqnarray}
\begin{claim}\label{claim:R_R_c_in_S_eps}
If a rate pair $(R, R_c) \in \mathbb{R}_{\geq 0}^2$ is $(\textcolor{black}{q_{X,Z},}q_{X,Y})$-achievable, then
$\forall \varepsilon\in (0,1/4), \ (R,R_c) \in \mathcal{S}_\varepsilon.$
\end{claim}
\begin{IEEEproof}
Let $(R, R_c)$ be $(\textcolor{black}{q_{X,Z},}q_{X,Y})$-achievable.
Fix $\varepsilon
\in (0,1/4)
.$
Then, there exists $n\in\mathbb{N}$ and a $(n,R, R_c)$ code inducing a joint distribution $P$ such that
\begin{IEEEeqnarray}{c}
\|\textcolor{black}{P_{X^n,Y^n} - q_{X,Y}^{\otimes n}}\|_{TV} \leq \varepsilon.\label{eq:in_converse_r_s_c_introducing_code_TV_performance}
\end{IEEEeqnarray}
Let $T$ be a uniform random variable over $[n],$
independent of all other variables.
Define $W\text{=}(M, J, T).$ 
We prove that $(R,R_c)\in\mathcal{S}_\varepsilon$ through Claims \ref{claim:X_T_Z_T_M_J_T_Y_T_satisfies_most_properties_in_D_eps}, \ref{claim:rate_bounds}, and \ref{claim:cardinality_bound_can_be_satisfied}.
\begin{claim}\label{claim:X_T_Z_T_M_J_T_Y_T_satisfies_most_properties_in_D_eps}
$P_{\textcolor{black}{X_T,Z_T},W,Y_T}$ satisfies all properties in the definition \eqref{eq:def_D_epsilon} of $\mathcal{D}_\varepsilon$ other than the cardinality bound.
\end{claim}
\begin{IEEEproof}
From \eqref{eq:in_converse_r_s_c_introducing_code_TV_performance} and Lemma \ref{lemma:TV_joint_to_TV_marginal}, we have
\begin{IEEEeqnarray}{c}
\forall t \in [n], \ \|P_{X_t,Y_t} - q_{X,Y}\|_{TV} \leq \varepsilon.\nonumber
\end{IEEEeqnarray}
Since $P_{X_T,Z_T}$ is an arithmetic average of the $P_{X_t,Y_t},$ then from the triangle inequality for the TVD, we have
\begin{IEEEeqnarray}{c}
\|P_{X_T,Y_T} - q_{X,Y}\|_{TV} \leq \varepsilon.\nonumber
\end{IEEEeqnarray}
\textcolor{black}{
Since $(X^n,Z^n) \sim q_{X,Z}^{\otimes n},$ the distribution of $(X_T,Z_T)$ is $q_{X,Z}$ and $(X_T,Z_T)$ is independent of $T.$
Since for any $t\in[n],$ we have the Markov chain relation $X_t-Z_t-(M,J,Y_t),$ and $T$ is independent of $(X^n,Z^n,M,J,Y^n),$ then we have the relation $X_T-(Z_T,T)-(M,J,Y_T).$ Since $(X_T,Z_T)$ is independent of $T,$ then we have $X_T-Z_T-T.$ Therefore, from the chain rule for mutual information, we have
\begin{IEEEeqnarray}{c}
X_T-Z_T-(M,J,T,Y_T).\label{eq:proved_first_half_of_desired_Markov_chain_relation}
\end{IEEEeqnarray}
}
Since for any $t\in[n],$ we have the Markov chain relation $Z_t-(M,J)-Y_t,$ and $T$ is independent of $(Z^n,M,J,Y^n),$ then we have $Z_T-(M,J,T)-Y_T.$
\textcolor{black}{
Combining this with \eqref{eq:proved_first_half_of_desired_Markov_chain_relation} yields the desired Markov chain relation $X_T-Z_T-(M,J,T)-Y_T.$
}
\end{IEEEproof}
\begin{claim}\label{claim:rate_bounds}
$(R,R_c)$ satisfies the inequalities in the definition \eqref{eq:def_S_epsilon} of $\mathcal{S}_\varepsilon$ for distribution $P_{X_T,Z_T,W,Y_T}.$
\end{claim}
\begin{IEEEproof}
We have
\begin{IEEEeqnarray}{rCl}
    nR &\geq& H(M)\nonumber\\
    &\geq& H(M|J)\nonumber \\*
    &\geq& \scalebox{1.0}{$I(M;Z^n | J)$} \nonumber \\
    &=& \scalebox{1.0}{$I(M, J ;Z^n)$} \label{eq:converse_E_D_marginal_R_using_J_indep} \\
    &=& \scalebox{1.0}{$\sum_{t=1}^n I(M, J ; Z_t | Z_{1:t-1})$} \nonumber \\ 
    &=& \scalebox{1.0}{$\sum_{t=1}^n I(M, J, Z_{1:t-1}; Z_t)$} \nonumber \\
    &\geq& \scalebox{1.0}{$\sum_{t=1}^n I(M, J ; Z_t)$} \nonumber \\
    &=& \scalebox{1.0}{$n I(M, J; Z_T | T)$}
    \IEEEeqnarraynumspace\label{eq:converse_E_D_marginal_lower_bound_R_t_to_T} \\  
    &=& \scalebox{1.0}{$n I(W ; Z_T),$}\label{eq:converse_E_D_marginal_lower_bound_R_T_indep}
\end{IEEEeqnarray}
where \eqref{eq:converse_E_D_marginal_R_using_J_indep} follows from the independence between the common randomness and the sources; and \eqref{eq:converse_E_D_marginal_lower_bound_R_t_to_T} and \eqref{eq:converse_E_D_marginal_lower_bound_R_T_indep} follow from the independence of $T$ and all other variables and from the fact that variables in $\{(Z_t)\}_{t \in [n]}$ are i.i.d. Moreover, we have
\begin{IEEEeqnarray}{rCl}
    n(R+R_c) &\geq& I(M,J;\textcolor{black}{X^n,} Y^n) \nonumber \\*
    &=& \sum_{t=1}^n I(M,J;\textcolor{black}{X_t,}Y_t|\textcolor{black}{X_{1:t-1},}Y_{1:t-1})\nonumber \\*
    &=& \sum_{t=1}^n \big[ I(M,J,\textcolor{black}{X_{1:t-1},}Y_{1:t-1} ;\textcolor{black}{X_t,} Y_t)
    \nonumber\\*
    &-& I(\textcolor{black}{X_{1:t-1},} Y_{1:t-1} ;\textcolor{black}{X_t,} Y_t) \big] 
    \nonumber\\*
    &\geq& -ng(\varepsilon) + \sum_{t=1}^n I(M,J;\textcolor{black}{X_t,}Y_t)
    \label{eq:in_converse_using_Cuff_lemma_1st_time} 
    \\*
    &=& nI(M,J;\textcolor{black}{X_T,}Y_T|T)
    -ng(\varepsilon)
    \label{eq:in_converse_turning_sum_into_T_bound_R_R_c}\\*
    &=& nI(T,M,J;\textcolor{black}{X_T, }Y_T)
    - nI(T;\textcolor{black}{X_T,}Y_T) - ng(\varepsilon)
    \nonumber
    \\*
    &\geq& nI(W;\textcolor{black}{X_T,}Y_T)
    -2ng(\varepsilon)
    ,\label{eq:in_converse_using_Cuff_lemma_2nd_time}
\end{IEEEeqnarray}where \eqref{eq:in_converse_using_Cuff_lemma_1st_time} and \eqref{eq:in_converse_using_Cuff_lemma_2nd_time} follow from \eqref{eq:in_converse_r_s_c_introducing_code_TV_performance}, the fact that $\varepsilon\in (0,1/4),$ and \cite[Lemma~VI.3]{2013PaulCuffDistributedChannelSynthesis}; and \eqref{eq:in_converse_turning_sum_into_T_bound_R_R_c} follows from the independence of $T$ from all other variables.
\end{IEEEproof}
\begin{claim}\label{claim:cardinality_bound_can_be_satisfied}
There exists a distribution $p_{\textcolor{black}{X,}Z,W,Y}$ under which $W$ satisfies the cardinality bound in the definition \eqref{eq:def_D_epsilon} of $\mathcal{D}_\varepsilon,$ the Markov chain relation $\textcolor{black}{X-}Z-W-Y,$ and
\begin{IEEEeqnarray}{c}
p_{\textcolor{black}{X,}Z,Y} \equiv P_{\textcolor{black}{X_T,}Z_T,Y_T}, \\* H_p(Z|W) = H_P(Z_T|W), \\* H_p(X,Y|W) = H_P(X_T,Y_T|W).
\end{IEEEeqnarray}
\end{claim}
\begin{IEEEproof}
For any finite alphabet $\mathcal{L},$ we denote by $\Delta_{\mathcal{L}}$ the set of probability distributions on $\mathcal{L}.$
Consider the following subset of $\mathbb{R}^{|\mathcal{Z}||\mathcal{Y}|}$
\begin{IEEEeqnarray}{c}
\mathcal{E} := \{(Q_Z(z)Q_Y(y))_{z\in\mathcal{Z},y\in\mathcal{Y}} | Q_Z \in \Delta_{\mathcal{Z}}, Q_Y \in \Delta_{\mathcal{Y}}\}.
\end{IEEEeqnarray}
This set lies inside the following hyperplane of $\mathbb{R}^{|\mathcal{Z}||\mathcal{Y}|}$
\begin{IEEEeqnarray}{c}
\mathcal{H} := \{(v_{z,y})_{z\in\mathcal{Z},y\in\mathcal{Y}} | \sum_{z\in\mathcal{Z},y\in\mathcal{Y}} v_{z,y}=1\},
\end{IEEEeqnarray}
\noindent which has dimension $|\mathcal{Z}||\mathcal{Y}|-1.$
\textcolor{black}{
Define the linear map
\begin{IEEEeqnarray}{c}
\phi: (v_{z,y})_{z\in\mathcal{Z},y\in\mathcal{Y}} \mapsto (q_{X|Z=z}(x)v_{z,y})_{x\in\mathcal{X},z\in\mathcal{Z},y\in\mathcal{Y}},
\end{IEEEeqnarray}
which is injective.
Then, $\phi(\mathcal{E})$ lies inside of the affine subspace $\phi(\mathcal{H}),$ which is of dimension at most $|\mathcal{Z}||\mathcal{Y}|-1.$
}
Define the following subset of $\textcolor{black}{\phi}(\mathcal{H})\times\mathbb{R}^2$
\begin{IEEEeqnarray}{rCl}
\IEEEeqnarraymulticol{3}{l}{
\Tilde{\mathcal{E}} := \{(\textcolor{black}{q_{X|Z=z}(x)}Q_Z(z)Q_Y(y))_{\textcolor{black}{x\in\mathcal{X},}z\in\mathcal{Z},y\in\mathcal{Y}}
}
\nonumber\\*
&\qquad& \cup (H_{Q_Z}(Z), H_{Q_Z Q_Y \textcolor{black}{q_{X|Z}}} (\textcolor{black}{X,}Y)) | Q_Z \in \Delta_{\mathcal{Z}}, Q_Y \in \Delta_{\mathcal{Y}}\},\nonumber
\end{IEEEeqnarray}where $\cup$ denotes the concatenation of vectors.
\textcolor{black}{
Set $\Tilde{\mathcal{E}}$ is compact and connected because it is the image of $\mathcal{E}$ by a continuous map, and
}
$\mathcal{E}$ is compact and connected.
Moreover, \textcolor{black}{$\Tilde{\mathcal{E}}$} lies in $\textcolor{black}{\phi}(\mathcal{H})\times\mathbb{R}^2,$ which is a strict affine subspace of $\textcolor{black}{\phi}(\mathbb{R}^{|\mathcal{Z}||\mathcal{Y}|})\times\mathbb{R}^2$ by injectivity of $\phi.$
Therefore, by the Caratheodory theorem for connected sets --- see the proof of \cite[Lemma~VI.1]{2013PaulCuffDistributedChannelSynthesis} --- any point in the convex hull of \textcolor{black}{$\Tilde{\mathcal{E}}$} can be written as a convex combination of $|\mathcal{Z}||\mathcal{Y}|+1$ elements of \textcolor{black}{$\Tilde{\mathcal{E}}.$}
From Claim \ref{claim:X_T_Z_T_M_J_T_Y_T_satisfies_most_properties_in_D_eps}, $P_{\textcolor{black}{X_T,}Z_T,W,Y_T}$ satisfies the Markov chain relation $\textcolor{black}{X_T-}Z_T-W-Y_T.$ Therefore,
\begin{IEEEeqnarray}{rCl}
P_{\textcolor{black}{X_T,}Z_T,Y_T}(x,z,y) &=& \sum_{w}P_{W}(w)\textcolor{black}{q_{X|Z=z}(x)}
\qquad\qquad\qquad
\nonumber\\*
&&\qquad\cdot P_{Z_T|W=w}(z)P_{Y_T|W=w}(y),
\nonumber\\*
H_P(Z_T|W) &=& \sum_w P_W(w) H(P_{Z_T|W=w}),
\nonumber
\end{IEEEeqnarray}
and
\begin{IEEEeqnarray}{c}
\scalebox{0.9}{$H_P(\textcolor{black}{X_T,}Y_T|W) =$}
\sum_w P_W(w)
\qquad\qquad\qquad\qquad\qquad\qquad\quad \
\nonumber\\*
\qquad\qquad\qquad\qquad
\cdot H(\sum_z P_{Z_T|W=w}(z) \textcolor{black}{q_{X|Z=z}} P_{Y_T|W=w}).
\nonumber
\end{IEEEeqnarray}
Then, the following vector
\begin{IEEEeqnarray}{c}
(P_{\textcolor{black}{X_T,}Z_T,Y_T}(\textcolor{black}{x,}z,y))_{\textcolor{black}{x\in\mathcal{X},}z\in\mathcal{Z},y\in\mathcal{Y}}
\nonumber\\*
\qquad\qquad\qquad\qquad
\cup(H_P(Z_T|W),H_P(\textcolor{black}{X_T,}Y_T|W))\nonumber
\end{IEEEeqnarray}
belongs to the convex hull of $\textcolor{black}{\Tilde{\mathcal{E}}}.$ Therefore, it is a convex combination of $|\mathcal{Z}||\mathcal{Y}|+1$ points of the $\textcolor{black}{\Tilde{\mathcal{E}}},$ which is equivalent to Claim \ref{claim:cardinality_bound_can_be_satisfied}.
\end{IEEEproof}
\vspace{5pt}

Let $p_{\textcolor{black}{X,}Z,W,Y}$ be a distribution such as in Claim \ref{claim:cardinality_bound_can_be_satisfied}. From Claims \ref{claim:X_T_Z_T_M_J_T_Y_T_satisfies_most_properties_in_D_eps} and \ref{claim:cardinality_bound_can_be_satisfied}, we have $p\in\mathcal{D}_\varepsilon.$ Combining this with Claim \ref{claim:rate_bounds} yields $(R,R_c) \in \mathcal{S}_\varepsilon.$ This being true for any $\varepsilon\in (0,1/4),$ Claim \ref{claim:R_R_c_in_S_eps} is proved.
\end{IEEEproof}
\vspace{5pt}

Similarly to \cite[Lemma~VI.5]{2013PaulCuffDistributedChannelSynthesis}, we have the following.
\begin{claim}\label{claim:S_is_cap_S_epsilon}
\begin{IEEEeqnarray}{c}
\bigcap_{\varepsilon\in(0,1/4)} \mathcal{S}_\varepsilon = \mathcal{S}^{(\textcolor{black}{r.c.s.})}.
\end{IEEEeqnarray}
\end{claim}
\begin{IEEEproof}
The proof is very similar to that of Claim \ref{claim:S_is_closed} (itself very similar to the proof of \cite[Lemma~VI.5]{2013PaulCuffDistributedChannelSynthesis}).
\end{IEEEproof}
\vspace{5pt}

Combining Claims \ref{claim:R_R_c_in_S_eps} and \ref{claim:S_is_cap_S_epsilon} yields $\mathcal{A}^{(\textcolor{black}{r.c.s.})} \subset \mathcal{S}^{(\textcolor{black}{r.c.s.})},$ as desired.


\section{Achievability direction of Theorem \ref{thm:remote_strong_coordination_region}}\label{app:achievability_thm_r_s_c}
This proof closely tracks \cite[Section~V]{2013PaulCuffDistributedChannelSynthesis}.
The notation is similar to the one in \cite{2023YassineGunduzISITRDPSideInformation}.
Here, we prove that $\mathcal{S}^{(\textcolor{black}{r.c.s.})} \subseteq \overline{\mathcal{A}}^{(\textcolor{black}{r.c.s.})}.$
To define joint distributions, we use, with some abuse of notation, $Q_{X,Y} :=\mu_X \cdot \nu_{Y|X{=}x},$ with $x$ a dummy variable: this defines $Q_{X,Y}$ as the composition of marginal distribution $\mu_X$ and conditional probability kernel $\nu.$

\subsection{Random codebook}\label{sec:where_we_fix_epsilon}
Let $(R,R_c)$ be a rate pair in $\mathcal{S}^{(\textcolor{black}{r.c.s.})}.$
Let $p_{\textcolor{black}{X,}Z,Y,W}$ be a corresponding distribution from the definition of $\mathcal{S
}^{(\textcolor{black}{r.c.s.})}.$ Then, we have
\begin{equation}\label{eq:introducing_R}
    R \geq I_p(Z;W)
\end{equation}
\begin{equation}\label{eq:introducing_R_c}
    R+R_c \geq I_p(\textcolor{black}{X,}Y;W)
\end{equation}
\noindent Fix some $\varepsilon>0.$
For every $n \geq 1,$ let $\mathcal{C}^{(n)}$ be a random codebook with $\lfloor 2^{n(R+\varepsilon)}\rfloor \times \lfloor 2^{nR_c}\rfloor$ i.i.d. codewords sampled from $p_{W}^{\otimes n}.$ The codewords are indexed by pairs $(m,j).$ We denote this random codebook distribution by $\mathbb{Q}_{\mathcal{C}^{(n)}}.$
Given a realization $c^{(n)},$ the codeowrd of index $(m,j)$ is denoted $w^n(c^{(n)},m,j).$

\subsection{Distribution $Q$ and soft-covering lemma}\label{sec:Q}
\begin{figure}[t!]
    \centering\includegraphics[width=\columnwidth]{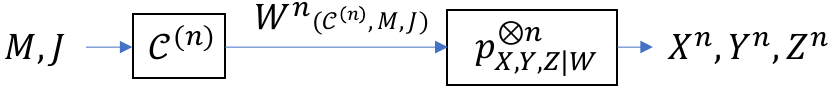}
    \caption{Graphical model for $Q.$ See Equations \eqref{eq:def_Q} and \eqref{eq:Q_averaged_over_codebook_gives_p}.}
    \label{fig:soft_covering_setup}
\end{figure}
For every positive integer $n,$ we define the distribution $Q,$ described in Figure \ref{fig:soft_covering_setup}, as follows:
\begin{IEEEeqnarray}{rCl}
Q_{\mathcal{C}^{(n)}, M,J,W^n,Z^n,Y^n\textcolor{black}{,X^n}}
&:=& \ \mathbb{Q}_{\mathcal{C}^{(n)}}
\cdot p^{\mathcal{U}}_{[2^{\scalebox{0.45}{$n(R{+}\varepsilon)$}}]}
\cdot p^{\mathcal{U}}_{[2^{nR_c}]}
\nonumber\\*
&\cdot& \mathbf{1}_{W^n{=}w^n(c^{(n)},m,j)}
\prod_{t=1}^n 
p_{\scalebox{0.55}{$Z|W\text{=}w_t$}}
\nonumber\\*
&\cdot&
\prod_{t=1}^n 
p_{\scalebox{0.55}{$Y|W\text{=}w_t$}}
\textcolor{black}{
\cdot
\prod_{t=1}^n 
p_{\scalebox{0.55}{$X|Z\text{=}x_t$}}.
}
\IEEEeqnarraynumspace\label{eq:def_Q}
\end{IEEEeqnarray}

\noindent By the definition of $\mathbb{Q}_{\mathcal{C}^{(n)}},$ we have $Q_{W^n} \equiv p_W^{\otimes n},$ and therefore, from \eqref{eq:def_Q} and the Markov chains in \eqref{eq:def_D_remote_strong_ccordination}, we have
\begin{IEEEeqnarray}{c}
Q_{W^n,Z^n,Y^n\textcolor{black}{,X^n}} \equiv p_{W,Z,Y\textcolor{black}{,X}}^{\otimes n}.\label{eq:Q_averaged_over_codebook_gives_p}
\end{IEEEeqnarray}
We use the \textit{soft covering lemma} 
\cite[Corollary~IV.1]{2013PaulCuffDistributedChannelSynthesis}.
A nearly direct application gives, from \eqref{eq:introducing_R} and \eqref{eq:introducing_R_c}:
\begin{claim}\label{claim:use_of_soft_covering_lemma}
\begin{IEEEeqnarray}{c}
\mathbb{E}_{\mathcal{C}^{(n)}}\big[\|Q_{\textcolor{black}{X^n,}Y^n|\mathcal{C}^{(n)}} - q_{\textcolor{black}{X,}Y}^{\otimes n}\|_{TV}\big] \underset{n \to \infty}{\longrightarrow} 0 \label{eq:TV_Q_X_Y}\\
\mathbb{E}_{\mathcal{C}^{(n)}}\big[\|Q_{J, Z^n|\mathcal{C}^{(n)}} - p^{\mathcal{U}}_{[2^{nR_c}]}q_{Z}^{\otimes n}\|_{TV}\big] \underset{n \to \infty}{\longrightarrow} 0.\IEEEeqnarraynumspace\label{eq:TV_Q_J_Z}
\end{IEEEeqnarray}
\end{claim}
\textcolor{black}{
The key difference with respect to the proof in \cite{2013PaulCuffDistributedChannelSynthesis} for standard channel synthesis is the following.
We can introduce the remote source $X$ via Lemma \ref{lemma:TV_same_channel} with $L=X^n,U=(J,Z^n),$ and $\Pi_{L|U}$ the memoryless channel $\prod q_{X|Z}$: from \eqref{eq:def_Q} and \eqref{eq:TV_Q_J_Z}, we get
\begin{IEEEeqnarray}{c}
\mathbb{E}_{\mathcal{C}^{(n)}}\big[\|Q_{J,Z^n,X^n|\mathcal{C}^{(n)}} - p^{\mathcal{U}}_{[2^{nR_c}]}q_{Z,X}^{\otimes n}\|_{TV}\big] \underset{n \to \infty}{\longrightarrow} 0.\IEEEeqnarraynumspace\label{eq:TV_Q_J_Z_X}
\end{IEEEeqnarray}
}
\begin{IEEEproof}[Proof of Claim \ref{claim:use_of_soft_covering_lemma}]
\begin{lemma}\label{lemma:soft_covering_basic}
\cite[Corollary~IV.1]{2013PaulCuffDistributedChannelSynthesis}
Let $\mathcal{W}$ be a finite alphabet and $\rho_W$ a distribution on the latter. Let $R$ be a non-negative real number and let $(k_n)_{n {\geq} 1}$ be a sequence of positive integers satisfying $k_n/2^{nR} \; \substack{\raisebox{-4pt}{$\to$} \\ \scalebox{0.5}{$n$$\to$$\infty$}} \; 1.$ For every positive integer $n,$ let $\mathcal{E}^{(n)}$ be a randomly generated collection of $k_n$ mutually independent sequences in $\mathcal{W}^n$ each drawn according to $\rho_W^{\otimes n}.$ Its distribution is denoted $\Gamma_{\mathcal{E}^{(n)}}.$ The sequences are indexed by some set $\mathcal{I}$ of size $k_n$ and in a realization $e^{(n)}$ of $\mathcal{E}^{(n)}$ the sequence with index $i$ is denoted by $w^n(e^{(n)}, i).$ A memoryless channel $(\rho_{U|W\text{=}w})_{w \in \mathcal{W}}$ induces an output distribution defined as
\begin{IEEEeqnarray}{c}
\text{$\Gamma$}
_{
\scalebox{0.6}{$
\mathcal{E}^{(n)},
I, W^n, U^n 
$}
}  
:=
\Gamma_{\mathcal{E}^{(n)}}
\cdot
p^{\mathcal{U}}_{[k_n]}
\cdot
\substack{\scalebox{1.2}{$\mathbf{1} \qquad \quad \quad $} \\ w^n = w^n(e^{(n)}, i)} 
\cdot
\prod_{t=1}^n \rho \substack{\scalebox{0.8}{$ (u_t) $} \\ \scalebox{0.55}{$U|W=w_t$} } \ .
\IEEEeqnarraynumspace
\end{IEEEeqnarray}
If $R > I_{\rho}(W;U),$ then
$$\mathbb{E}_{\mathcal{E}^{(n)}}\big[\|\Gamma_{U^n|\mathcal{E}^{(n)}} - \rho_{U}^{\otimes n}\|_{TV}\big] \underset{n \to \infty}{\longrightarrow} 0,$$
\noindent where $\rho_U$ is the marginal of $\rho_{W,U} = \rho_W \cdot \rho_{U|W}.$
\end{lemma}

We first use Lemma \ref{lemma:soft_covering_basic} with $\rho = p,$ $\mathcal{E}^{(n)}=\mathcal{C}^{(n)},$ $k_n = \lfloor 2^{n(R+\varepsilon)}\rfloor \times \lfloor 2^{nR_c}\rfloor,$ $I=(M,J),$ $U=(\textcolor{black}{X,}Y)$ and $R$ replaced by $R+\varepsilon+R_c.$
The corresponding output distribution is exactly $Q_{\mathcal{C}^{(n)},M,J, W^n,\textcolor{black}{X^n,} Y^n},$ by definition \eqref{eq:def_Q} of the latter. Indeed, from \eqref{eq:def_Q} distribution $Q$ satisfies the Markov chain relation $(\mathcal{C}^{(n)},M,J)-W^n-(\textcolor{black}{X^n,}Y^n),$ and from \eqref{eq:Q_averaged_over_codebook_gives_p} we have $Q_{W^n,\textcolor{black}{X^n,} Y^n} \equiv p_{W,\textcolor{black}{X,}Y}^{\otimes n}.$ 
\todo[inline]{Indeed, the Markov chain $(M,J)-W^n-Y^n$ implies that we can write $Q$ as $Q_{Y^n,W^n} \cdot Q_{\mathcal{C}^{(n)},M,J|W^n=w^n}.$}
Thus, from \eqref{eq:introducing_R_c} and Lemma \ref{lemma:soft_covering_basic} together with the coordination property $p_{\textcolor{black}{X,}Y} \equiv q_{\textcolor{black}{X,}Y}$ from \eqref{eq:def_D_remote_strong_ccordination} we get \eqref{eq:TV_Q_X_Y}.\\

Second, we use Lemma \ref{lemma:soft_covering_basic} as follows.
For every $n\in\mathbb{N},$
we set $\mathcal{E}^{(n)}$ to be the sub-codebook of $\mathcal{C}^{(n)}$ corresponding to $j=1,$ denoted $\mathcal{C}^{(n)}_1,$ and set $\rho = p,$ $k_n = \lfloor 2^{n(R+\varepsilon)}\rfloor.$
We further set $I=(M,1),$ $U=Z,$ and replace $R$ in the lemma by $R+\varepsilon.$
The corresponding output distribution is exactly $Q_{\mathcal{C}^{(n)}_1,M,J, W^n, Z^n |J{=}1},$ by definition \eqref{eq:def_Q} of the latter. 
Indeed, from \eqref{eq:def_Q} distribution $Q$ satisfies $(\mathcal{C}^{(n)},M,J)-W^n-Z^n,$ and from \eqref{eq:Q_averaged_over_codebook_gives_p} we have $Q_{W^n, Z^n} \equiv p_{W,Z}^{\otimes n}.$ 
From \eqref{def:code_remote_strong_coordination}, we have $p_Z \equiv q_Z.$ 
Thus, from \eqref{eq:introducing_R} and Lemma \ref{lemma:soft_covering_basic} we have
\begin{equation}\label{eq:use_of_soft_covering_TV_Q_X_Z}
    \mathbb{E}_{\mathcal{C}^{(n)}_1}\big[\|Q_{Z^n|\mathcal{C}^{(n)}_1, J=1} - q_{Z}^{\otimes n}\|_{TV}\big] \underset{n \to \infty}{\longrightarrow} 0.
\end{equation}
From \eqref{eq:def_Q}, we have $\forall j, \ Q_{Z^n|\mathcal{C}^{(n)}_j, J=j} = \psi(\mathcal{C}^{(n)}_j),$ where
\begin{IEEEeqnarray}{c}
\psi:
\{w^n(m)\}_{m\in[k_n]} \in (\mathcal{W}^n)^{k_n} \mapsto \dfrac{1}{k_n} \sum_{m=1}^{k_n} \prod_{t=1}^n p_{Z|W{=}w_t(m)}\nonumber
\end{IEEEeqnarray}
is a map which is defined independently of any index $j.$ Therefore, $Q_{Z^n|\mathcal{C}^{(n)},J} \equiv Q_{Z^n|\mathcal{C}^{(n)}_J,J}.$
\noindent Moreover, since from \eqref{eq:def_Q}  we have $Q_{\mathcal{C}^{(n)}, J} \equiv \mathbb{Q}_{\mathcal{C}^{(n)}}p^{\mathcal{U}}_{[2^{nR_c}]},$ then, from Lemma \ref{lemma:get_expectation_out_of_TV},
\begin{IEEEeqnarray}{c}
\|Q_{\mathcal{C}^{(n)},J, Z^n} - \mathbb{Q}_{\mathcal{C}^{(n)}} p^{\mathcal{U}}_{[2^{nR_c}]} p_{Z}^{\otimes n}\|_{TV}\nonumber\\*
=\sum_{j=1}^{\lfloor 2^{nR_c} \rfloor} \tfrac{\text{\normalsize 1}}{\lfloor 2^{nR_c} \rfloor} \mathbb{E}_{\mathcal{C}^{(n)}_j}\big[\|Q_{Z^n|\mathcal{C}^{(n)}_j, J=j} - q_{Z}^{\otimes n}\|_{TV}\big].
\label{eq:average_over_j_of_identical_expectations}
\end{IEEEeqnarray}
Moreover,
by construction of $\mathcal{C}^{(n)},$ the distribution of sub-codebook $\mathcal{C}^{(n)}_j=(W^n(\mathcal{C}^{(n)}, m,j))_{m}$ is independent of $j.$ Hence, this is also the case for $\psi(\mathcal{C}^{(n)}_j),$ and all expectations in
\eqref{eq:average_over_j_of_identical_expectations}
are identical. Using \eqref{eq:use_of_soft_covering_TV_Q_X_Z} and Lemma \ref{lemma:get_expectation_out_of_TV} we get \eqref{eq:TV_Q_J_Z}.
\end{IEEEproof}


\subsection{Choosing a codebook}
We know that $L_1$ convergence implies convergence in probability. Therefore,
we get
convergence in probability from \eqref{eq:TV_Q_X_Y} and \eqref{eq:TV_Q_J_Z_X}. Then, for a certain $N_0,$ for every integer $n\geq N_0$ there is a codebook $c_*^{(n)}$ such that
\begin{IEEEeqnarray}{c}
    \scalebox{0.9}{$\|Q_{\textcolor{black}{X^n,}Y^n|\mathcal{C}^{(n)}=c_*^{(n)}} - q_{\textcolor{black}{X,}Y}^{\otimes n}\|_{TV} \leq \varepsilon, $}\label{eq:TV_Q_X_Y_fixed_codebook} \IEEEeqnarraynumspace\\
    \scalebox{0.9}{$\|Q_{J, Z^n\textcolor{black}{, X^n}|\mathcal{C}^{(n)}=c_*^{(n)}} - p^{\mathcal{U}}_{[2^{nR_c}]}q_{Z\textcolor{black}{,X}}^{\otimes n}\|_{TV} \leq \varepsilon. $}\label{eq:TV_Q_J_Z_X_fixed_codebook} \IEEEeqnarraynumspace
\end{IEEEeqnarray}

\subsection{Construction of a code}\label{subsec:defining_P}
In this subsection, we show how the distribution 
\scalebox{0.9}{$Q_{|\mathcal{C}^{(n)}{=}c^{(n)}_*}$} can be modified
to lead to a code,
with a negligible effect on the coordination performance.
We define the distribution $P$ which constitutes a $(n,R+\varepsilon,R_c)$ code. It differs from $Q$ in having the correct marginal for $(\textcolor{black}{X^n,}Z^n,J){:}$ let
\begin{IEEEeqnarray}{rCl}
P_{J,\textcolor{black}{X^n,}Z^n,M,Y^n}
&:=&
p^{\mathcal{U}}_{[2^{nR_c}]}
\cdot
q_{\scalebox{0.7}{$\textcolor{black}{X,} Z$}}^{\otimes n}
\cdot
Q_{\scalebox{0.7}{$M|\mathcal{C}^{(n)}{=}c^{(n)}_*, J\text{=}j, Z^n\text{=}z^n$}}
\nonumber\\*
&\cdot&
Q_{\scalebox{0.7}{$Y^n|\mathcal{C}^{(n)}{=}c^{(n)}_*, M{=}m, J{=}j$}}
,\label{eq:long_def_P}
\IEEEeqnarraynumspace
\end{IEEEeqnarray}
which from \eqref{eq:def_Q} can be written as
\begin{IEEEeqnarray}{rCl}
P_{J,\textcolor{black}{X^n,}Z^n,M,Y^n}
&=&
p^{\mathcal{U}}_{[2^{nR_c}]}
\cdot q_{\scalebox{0.7}{$\textcolor{black}{X,} Z$}}^{\otimes n}
\cdot Q_{\scalebox{0.7}{$M,Y^n|\mathcal{C}^{(n)}{=}c^{(n)}_*, J\text{=}j, Z^n\text{=}z^n$}}
.\label{eq:short_def_P}
\IEEEeqnarraynumspace
\end{IEEEeqnarray}
From \eqref{eq:long_def_P} distribution $P$ satisfies
\begin{IEEEeqnarray}{c}
\textcolor{black}{X^n-}Z^n-(M,J)-Y^n.\label{eq:long_Markov_for_P}
\end{IEEEeqnarray}
The alphabet of $M$ is $[2^{n(R+\varepsilon)}]$ throughout this proof by construction of the random codebook $\mathcal{C}^{(n)}.$
Therefore, from \eqref{eq:short_def_P}-\eqref{eq:long_Markov_for_P}
and Lemma \ref{lemma:conditions_for_P_to_define_a_code},
distribution $P$ defines a $(n,R+\varepsilon,R_c)$ code.
From \eqref{eq:def_Q}, \eqref{eq:short_def_P}, and Lemma \ref{lemma:TV_same_channel}, comparing $P$ with $Q_{|\mathcal{C}^{(n)}{=}c^{(n)}_*}$ reduces to comparing marginals, i.e., to \eqref{eq:TV_Q_J_Z_X_fixed_codebook}:
\begin{IEEEeqnarray}{rCl}
\IEEEeqnarraymulticol{3}{l}{
\big\|P_{\scalebox{0.6}{$J,\textcolor{black}{X^n,}Z^n,M,Y^n$}} - Q_{\scalebox{0.6}{$J,\textcolor{black}{X^n,}Z^n,M,Y^n|\mathcal{C}^{(n)}{=}c^{(n)}_*$}}\big\|_{TV}
}\nonumber\\*
\quad & = & \big\|P_{J, \textcolor{black}{X^n,} Z^n} - Q_{J, \textcolor{black}{X^n,} Z^n|\mathcal{C}^{(n)}{=}c^{(n)}_*}\big\|_{TV}\leq \varepsilon. \label{eq:TV_P_Q}
\end{IEEEeqnarray}
Therefore, from Lemma \ref{lemma:TV_joint_to_TV_marginal} with $U=(\textcolor{black}{X^n,} Y^n),$ \eqref{eq:TV_Q_X_Y_fixed_codebook}, and the triangle inequality, we get
\begin{IEEEeqnarray}{c}
    \big\|P_{\textcolor{black}{X^n,} Y^n} - 
    q_{\textcolor{black}{X,}Y}^{\otimes n}
    \big\|_{TV} \leq 2\varepsilon. \label{eq:TV_sur_X_Y_P_Q}
\end{IEEEeqnarray}
This proves that $(R+\varepsilon, R_c)$ is $(\textcolor{black}{q_{X,Z},}q_{X,Y})$-achievable.
This is true for every $\varepsilon>0.$ Therefore, we get $(R, R_c) \in \overline{\mathcal{A}}^{(\textcolor{black}{r.c.s.})}.$
This being true for any $(R,R_c)\in\mathcal{S}^{(\textcolor{black}{r.c.s.})}$, we have $\mathcal{S}^{(\textcolor{black}{r.c.s.})} \subseteq \overline{\mathcal{A}}^{(\textcolor{black}{r.c.s.})},$ as desired.


\section{Proof of Proposition \ref{prop:suboptimality_of_of_direct_strong_coordination}}\label{app:proof_binary_strict_suboptimality_of_direct_strong_coordination}
Since the notion of compatibility with $(q_{X,Z},q_{X,Y})$ consists of linear conditions, the set $\cup_{q_{X,Z,Y}}\mathcal{D}^{(d.c.s.)}(q_{X,Z,Y})$ is closed. Since it is a subset of a finite-dimensional real vector space, then it is compact. Therefore, by continuity of the mutual information over finite alphabets, and due to non-strict inequalities, $\cup_{q_{X,Z,Y}}\mathcal{S}^{(d.c.s.)}(q_{X,Z,Y})$ is closed. Similarly, $\mathcal{S}^{(r.c.s.)}$ is also closed. This justifies the existence of the minima in \eqref{eq:statement_suboptimality_of_d_s_c}.
\begin{claim}
It is sufficient to prove \eqref{eq:statement_suboptimality_of_d_s_c} for $R_c=0.$
\end{claim}
\begin{IEEEproof}
Assume that \eqref{eq:statement_suboptimality_of_d_s_c} holds for $R_c=0.$
Denote by $r$ the corresponding positive difference between both sides of \eqref{eq:statement_suboptimality_of_d_s_c}, and by $R_0$ the rate achieving the minimum in the right hand side.
Fix $R_c \in (0,r).$
For any $q_{X,Z,Y}$ and any $R\in\mathbb{R}_{\geq 0}$ such that $(R,R_c)\in \mathcal{S}^{(d.c.s.)}(q_{X,Z,Y}),$ we have
from the definition of $\mathcal{S}^{(d.c.s.)}(q_{X,Z,Y})$ (Eq. \ref{eq:def_S_direct_strong_coordination})
that
$(R{+}R_c,0)\in\mathcal{S}^{(d.c.s.)}(q_{X,Z,Y}),$
hence $R+R_c \geq R_0+r > R_0+R_c.$
Then, \eqref{eq:statement_suboptimality_of_d_s_c} follows from the fact that $(R_0,R_c)\in \mathcal{S}^{(r.c.s.)}$ (since $(R_0,0)\in \mathcal{S}^{(r.c.s.)}$).
\end{IEEEproof}

We consider $R_c=0$ in the remainder of Appendix \ref{app:proof_binary_strict_suboptimality_of_direct_strong_coordination}. Fix $R$ as the rate attaining the first minimum, and consider a distribution $q_{X,Z,Y}$ compatible with $(q_{X,Z},q_{X,Y})$ such that $(R,0)\in\mathcal{S}^{(d.c.s.)}(q_{X,Z,Y}).$ Then, there exists a distribution $p_{X,Z,W,Y}\in\mathcal{D}^{(d.c.s.)}(q_{X,Z,Y})$ such that $R \geq I_p(Z,Y;W).$ By minimality of $R,$ we have $R=I_p(Z,Y;W).$ Therefore, $R\geq C_p(Z,Y),$ where $C_p(Z,Y)$ denotes the Wyner common information between $Z$ and $Y,$ defined as
\begin{IEEEeqnarray}{c}
C_p(Z,Y) := \min_{W \text{ s.t. } Z-W-Y} I(Z,Y;W).
\end{IEEEeqnarray}
We only have the following five cases.\\

Case 1: $R>C_p(Z,Y).$\\
Let $\Tilde{R}\in(C_p(Z,Y),R).$
Then, by definition of the Wyner common information, there exists $p_{\Tilde{W}|Z,Y}$ such that $p_{Z,Y}\cdot p_{\Tilde{W}|Z,Y}$ satisfies the Markov chain relation $Z-\Tilde{W}-Y$ and $I_p(Z,Y;\Tilde{W})\leq \Tilde{R}.$
Therefore, we have $(\Tilde{R},0)$ belongs to $\mathcal{S}^{(d.c.s.)}(q_{X,Z,Y}),$ by definition of the latter.
By construction, we have $\Tilde{R}<R.$ This contradicts the minimality of $R.$\\

Case 2: $R=C_p(Z,Y)$ and $q_{Z,Y}(Y\neq Z)=0.$\\
Then, $q_{X,Y}(X\neq Y)=q_{X,Z}(X\neq Z),$ which contradicts the assumptions.\\

Case 3: $R=C_p(Z,Y)$ and $q_{Z,Y}(Y\neq Z)=1.$\\
Then, $q_{X,Y}(X\neq Y)=1-q_{X,Z}(X\neq Z)>1/2,$ which contradicts the assumptions.\\

Case 4: $R=C_p(Z,Y)$ and $q_{Z,Y}(Y\neq Z)\in[1/2,1).$\\
Define $\theta=q_{Z,Y}(Y\neq Z)$ and $\tau=q_{X,Z}(X\neq Z).$
Then,
\begin{IEEEeqnarray}{rCl}
q_{X,Y}(X\neq Y)&=&\tau(1-\theta)+(1-\tau)\theta
\nonumber\\*
&=& 1/2+(\theta-1/2)(1-2\tau)\geq1/2.
\end{IEEEeqnarray}
This contradicts the assumptions.\\

Case 5: $R=C_p(Z,Y)$ and $q_{Z,Y}(Y\neq Z)\in(0,1/2).$\\
Define $\theta = q_{Z,Y}(Y\neq Z).$
Since $q_Z\equiv q_Y \equiv Ber(1/2),$ then $(Z,Y)$ is a \textit{doubly binary symmetric source of parameter $\theta$} \cite[Example~10.1]{2011BookElGamalKimNetworkInformationTheory}, that is $q_{Z,Y}(0,1)=q_{Z,Y}(1,0)=\theta/2$ and $q_{Z,Y}(0,0)=q_{Z,Y}(1,1)=(1-\theta)/2.$
This is abbreviated as $q_{Z,Y} \equiv DSBS(\theta).$
Therefore,
as shown in \cite[Section~3]{1975WynerCommonInformationAndSoftCoveringLemma}, one obtains $p_{Z,Y}\equiv q_{Z,Y}$ and $I_p(Z,Y;\Tilde{W})=C_p(Y,Z)$ when choosing $p_{Z,\Tilde{W},Y}$ of the form $q_Z\cdot p_{\Tilde{W}|Z}\cdot p_{Y|Z}$ with $p_{\Tilde{W}|Z}$ and $p_{Y|\Tilde{W}}$ identical BSC($\Tilde{\theta}$), where
\begin{IEEEeqnarray}{c}
\Tilde{\theta} = \dfrac{1}{2} - \dfrac{1}{2}\sqrt{1-2\theta}.
\end{IEEEeqnarray}
Define $p_{X,Z,\Tilde{W},Y}:=p_{X|Z} \cdot p_{Z,\Tilde{W},Y}$ and
\begin{IEEEeqnarray}{c}
\Tilde{R} := \max(I_p(Z;\Tilde{W}), I_p(X,Y;\Tilde{W})).
\end{IEEEeqnarray}
Then $(\Tilde{R},0)$ belongs to $\mathcal{S}^{(r.c.s.)},$ by definition of the latter.
It remains to show that $\Tilde{R}<R.$
Let $h(\cdot)$ denote the binary entropy function.
We have
\begin{IEEEeqnarray}{rCl}
R-I_p(Z;\Tilde{W}) &=& I_p(Z,Y;\Tilde{W}) - I_p(Z;\Tilde{W})
\nonumber\\*
&=& I_p(Y;\Tilde{W}|Z)
\nonumber\\*
&=& I_p(Y;\Tilde{W},Z) - I_p(Z;Y)
\nonumber\\*
&=& I_p(Y;\Tilde{W})-I_p(Z;Y)
\label{eq:decomposition_of_terms_like_Wyner_common_info}\\*
&=& 1-h(\Tilde{\theta}) -1 + h(\theta)>0,
\end{IEEEeqnarray}
where the last line follows from the fact that $\Tilde{\theta}<\theta<1/2,$ which follows from the fact that $\sqrt{u}>u$ for $u\in (0,1).$
It remains to show that $R>I_p(X,Y;\Tilde{W}).$
By construction,
\begin{IEEEeqnarray}{c}
p_{Z,\Tilde{W}}\equiv p_{\Tilde{W},Y}\equiv DSBS(\Tilde{\theta}).
\end{IEEEeqnarray}
Moreover, with $\tau:=q_{X,Z}(X\neq Z),$ we have
\begin{IEEEeqnarray}{rCl}
p_{X,\Tilde{W}} &\equiv& DSBS((1-\tau)\Tilde{\theta} + \tau(1-\Tilde{\theta}))\nonumber\\*
&\equiv& DSBS(1/2 - (1/2-\tau)\sqrt{1-2\theta}),
\IEEEeqnarraynumspace\\*
p_{X,Y} &\equiv& DSBS((1-\tau)\theta + \tau(1-\theta))
\nonumber\\*
&\equiv& DSBS(\tau+(1-2\tau)\theta).\nonumber
\end{IEEEeqnarray}
Then,
\begin{IEEEeqnarray}{rCl}
R - I_p(X,Y;\Tilde{W}) &=&
I_p(Z,Y;\Tilde{W}) - I_p(X,Y;\Tilde{W})
\nonumber\\*
&=& I_p(Z;\Tilde{W}) + I_p(Y;\Tilde{W}) - I_p(Z;Y)
\nonumber\\*
&-& I_p(X;\Tilde{W}) - I_p(Y;\Tilde{W}) + I_p(X;Y)
\label{eq:using_decomposition_of_terms_like_Wyner_common_info}\\*
&=& 1-h(1/2 - (1/2)\sqrt{1-2\theta}) -1 + h(\theta)\nonumber\\*
&-& 1 + h(1/2 - (1/2-\tau)\sqrt{1-2\theta})\nonumber\\*
&+& 1-h(\tau+(1-2\tau)\theta),\label{eq:closed_form_gap_between_R_and_Tilde_R}
\end{IEEEeqnarray}
where each line in \eqref{eq:using_decomposition_of_terms_like_Wyner_common_info} follows from the chain of equalities leading to \eqref{eq:decomposition_of_terms_like_Wyner_common_info}.
A plot of the right hand side of \eqref{eq:closed_form_gap_between_R_and_Tilde_R} is provided in Figure \ref{fig:plot_gap_optimal_rate_remote_coordination_vs_analog}.
To conclude the proof, we use the following lemma.
\begin{lemma}\label{lemma:Bodo_concave}
Consider reals $\alpha,\beta$ such that $\alpha<\beta,$ and a concave increasing function $f:(\alpha,\beta)\to\mathbb{R}.$ Consider $a,b,c,d$ in $(\alpha,\beta)$ such that
\begin{IEEEeqnarray}{c}
a < \min(b,c), \ d>\max(b,c), \ b+c > a+d.\label{eq:conditions_lemma_Bodo_concave}
\end{IEEEeqnarray}
Then, $-f(a)+f(b)+f(c)-f(d)>0.$
\end{lemma}
\begin{IEEEproof}
Denote $b'=\min(b,c)$ and $c'=\max(b,c).$
Quantity $-f(a)+f(b)+f(c)-f(d)$ rewrites as
\begin{IEEEeqnarray}{c}
(b'-a)\dfrac{f(b')-f(a)}{b'-a} - (d-c')\dfrac{f(d)-f(c')}{d-c'}.
\end{IEEEeqnarray}
By assumption, we have that $b'-a>d-c'.$ By concavity of $f,$ and since $\max(a,b')\leq\min(c',d),$ the slope between $a$ and $b'$ is greater or equal to the slope between $c'$ and $d.$ These slopes are (strictly) positive because $f$ is increasing, which concludes.
\end{IEEEproof}

It remains to check that we can apply the lemma, with $f=h,$ to the right hand side of \eqref{eq:closed_form_gap_between_R_and_Tilde_R}. First, $h$ is concave and increasing on $(0,1/2).$ It is straightforward to check that the arguments of $h$ in \eqref{eq:closed_form_gap_between_R_and_Tilde_R} lie in $(0,1/2)$, and that the first two conditions in \eqref{eq:conditions_lemma_Bodo_concave} hold. Moreover,
\begin{IEEEeqnarray}{rCl}
\IEEEeqnarraymulticol{3}{l}{
\theta+[1/2-(1/2-\tau)\sqrt{1-2\theta}]
}\nonumber\\*
&-& [1/2-1/2\sqrt{1-2\theta}] - [\tau+(1-2\tau)\theta]\nonumber\\*
&=&\tau\sqrt{1-2\theta}-\tau(1-2\theta)>0.
\end{IEEEeqnarray}
Thus, the third condition in \eqref{eq:conditions_lemma_Bodo_concave} holds, and Lemma \ref{lemma:Bodo_concave} can be applied.
This concludes the proof in Case 5, which concludes the proof of Proposition \ref{prop:suboptimality_of_of_direct_strong_coordination}.

\end{document}